\documentclass[utf8]{FrontiersinHarvard} 

\usepackage{graphicx,natbib,booktabs}
\usepackage{multicol,multirow}
\usepackage{amsmath,amssymb,amsfonts}
\usepackage{mathrsfs}
\usepackage{amsthm}
\usepackage{rotating}
\usepackage{appendix}
\usepackage{ifpdf}
\usepackage[T1]{fontenc}
\usepackage{newtxtext}
\usepackage{newtxmath}
\usepackage{textcomp}
\usepackage{xcolor}
\usepackage{lipsum}
\usepackage{tikz}
\usetikzlibrary{arrows.meta,positioning,calc,shapes,decorations.pathmorphing,fit,backgrounds}
\theoremstyle{plain}
\newtheorem{theorem}{Theorem}[section]
\newtheorem{proposition}[theorem]{Proposition}

\newtheorem{corollary}[theorem]{Corollary}
\theoremstyle{definition}
\newtheorem{definition}[theorem]{Definition}

\theoremstyle{remark}
\newtheorem{remark}[theorem]{Remark}

\newcommand{\Heven}{\mathcal{H}_{\text{even}}}
\newcommand{\Hodd}{\mathcal{H}_{\text{odd}}}

\def\keyFont{\fontsize{8}{11}\helveticabold }
\def\firstAuthorLast{Li} 
\def\Authors{Xin Li\,$^{1,*}$ }
\def\Address{$^{1}$Department of Computer Science, University at Albany, Albany, NY 12222 }
\def\corrAuthor{Corresponding Author}

\def\corrEmail{xli48@albany.edu}

\begin{document}
\onecolumn
\firstpage{1}

\title {The Homological Brain: Parity Principle and Amortized Inference} 

\author[\firstAuthorLast ]{\Authors} 
\address{} 
\correspondance{} 

\extraAuth{}

\maketitle

\begin{abstract}

We propose that biological intelligence is fruitfully understood as \emph{structure-preserving compression}: the brain does not minimize a per-sample loss, but extracts and preserves the topological invariants of experience through which future inference can be amortized.
Biological intelligence emerges from substrates that are slow, noisy, and
energetically constrained, yet it performs rapid and coherent inference in
open-ended environments. Classical computational theories, built around
vector-space transformations and instantaneous error minimization, struggle to
reconcile the slow timescale of synaptic plasticity with the fast timescale of
perceptual synthesis. We propose a unifying framework based on algebraic
topology, the \textbf{Homological Brain}, in which neural computation is
understood as the construction and navigation of topological structure. Central
to this view is the \textbf{Parity Principle}, a hypothesized homological partition between
even-dimensional \emph{scaffolds} encoding stable content ($\Phi$) and
odd-dimensional \emph{flows} encoding dynamic context ($\Psi$). Transient
contextual flows are resolved through a three-stage \textbf{topological trinity
transformation}: \emph{Search} (open-chain exploration), \emph{Closure}
(topological cycle formation), and \emph{Condensation} (collapse of validated
flows into new scaffolds). This process is \emph{structurally analogous} to the space-time
tradeoff captured by Savitch's theorem: costly recursive exploration is reorganized into efficient
memoized navigation over a learned manifold. We operationalize the framework with an
explicit pipeline that maps spike trains to chain complexes and validate the two central claims
in controlled simulations: (i) even- and odd-dimensional Betti numbers anti-correlate
across simulated wake/sleep cycles, consistent with parity alternation, and (ii) an agent
that condenses closed cycles into reusable scaffold vertices achieves a $\sim$3-4$\times$
inference-cost speedup over ab-initio search on a structured planning task, with a
scale-dependent advantage over flat-memory caching that emerges as the environment's
combinatorial complexity grows, consistent with amortization.
Taken together, this perspective unifies the Wake-Sleep cycle, episodic-to-semantic
consolidation, and dual-process theories (System~1 vs.\ System~2), framing the
brain as a homology engine that converts high-entropy sensory flux into low-entropy, invariant cognitive structure
through topologically-constrained reuse of prior inference.

 \keyFont{ \section{Keywords:} Homological brain, parity principle, scaffold-flow memory model, Memory-amortized inference (MAI), topological trinity transformation, topological wake-sleep algorithm} 
\end{abstract}

\section{Introduction}

Biological brains achieve a level of flexible, data-efficient intelligence that remains unmatched by current artificial systems \citep{LeCun2015}. Neurons operate at millisecond timescales, while synaptic updates necessary for learning unfold over minutes to hours, and total energy budgets are severely constrained (approx. 20W in humans) \citep{mead2002neuromorphic}. From these spatiotemporal limitations emerges an architecture capable of rapid perceptual inference, stable memory formation, and adaptive action selection \citep{friston2010free}. A central challenge in theoretical neuroscience is therefore to explain how slow, local biophysical processes give rise to fast, global coherence in cognition \citep{dayan2005theoretical}. How does the brain bridge the gap between the chaotic flux of sensory stimuli and the stable manifolds of conscious perception \citep{mountcastle1998perceptual} with such a tight energy budget?

Classical computational metaphors, from feedforward feature extraction to iterative error minimization (e.g., backpropagation \citep{rumelhart1985learning}), offer partial explanations. However, they lack a unifying principle that links neural dynamics (activity) \citep{gerstner2014neuronal}, structural organization (connectivity) \citep{bullmore2012economy}, and learning (plasticity) across scales \citep{hebb2005organization}. While we understand how single neurons integrate potentials, and how vast networks might classify images, we lack a rigorous geometric description of how information is \emph{conserved} and \emph{structured} during transmission \citep{amari2016information}. As argued by Carver Mead in \citep{mead2012analog}, ``our ability to realize simple neural functions is strictly limited by our understanding of their organizing principles.''

In this paper, we argue that such a principle must be \emph{topological} \citep{curto2025topological,giusti_clique_2015,sizemore2019importance}. The closure identity $\partial^{2}=0$ is an algebraic tautology once a chain complex is given; we do \emph{not} claim it acts as an additional physical law. Rather, we adopt it as an \emph{organizing modeling constraint}: the hypothesis we investigate is that successful neural computation admits a description as a chain complex in which observed activity patterns correspond to cycles (closed inferences) and errors to non-closed chains. Under this hypothesis, $\partial^{2}=0$ plays the role that energy conservation plays in physics-based modeling: a bookkeeping identity that, when imposed, disciplines the structure of admissible dynamics \citep{wheeler1990information,edelsbrunner2022computational}. We formalize this modeling stance through the \textbf{Parity Principle}, a \emph{hypothesis} that the scaffold/flow (even/odd homology) partition is a useful functional dichotomy for describing neural coding, and the mechanism of \textbf{Amortized Inference} \citep{gershman2014amortized}, which treats the conversion of recursive search into memoized lookup. Note that both parity alternation and memory amortization have classical precedents: 1) the constructive proof of Urysohn's lemma \citep{munkres2025elements} builds a continuous separating function by iteratively inserting new open sets at \emph{odd}-numerator dyadic levels between the \emph{even}-numerator sets of the prior level, converting topological separation into a metric function through a parity-alternating refinement; 2) memory-amortized inference (MAI) is \emph{analogous} to the space-time tradeoff captured by Savitch's theorem in computational complexity theory \citep{savitch1970relationships,bellman1966dynamic}. Throughout, we distinguish analogies (formal structural parallels) from established empirical facts; operational definitions and falsifiable predictions are deferred to Section~\ref{sec:empirical}, where we make them explicit and test them in simulation.

We find it useful to frame the resulting picture as a \emph{metric-topological factorization} (MTF) of the brain's representational problem, building on the long-standing distinction in cognitive neuroscience between an associative topological index in the medial temporal system and a continuous, amortized metric store in neocortex \citep{mcclelland1995there,oreilly2014complementary}. In this framing, the hippocampus and associated structures act as a topological index, a sparse, stable inventory of which experiences have closed into coherent cycles, while the neocortex stores the amortized transport plan that maps new contexts onto those previously-closed structures. Intelligence, in this view, is compression that preserves the homotopy type of experience: the system is not minimizing a per-sample loss but factoring out the topological invariants along which subsequent inference is cheap. The parity principle specifies \emph{how} this factorization is encoded, in the even versus odd dimensions of an underlying chain complex, and MAI specifies \emph{how} the factorization is computed, through cycles of search, closure, and condensation. The empirical signatures we test in Section~\ref{sec:empirical} are direct consequences of this factorization view.

\section{Foundation: Parity Principle and Amortized Inference}

\subsection{The Parity Principle: Computation as Topological Closure}

A chain complex encodes two complementary aspects of computation:
the \emph{dot-cycle dichotomy}, analogous to the zero-one dichotomy in digital logic.
Dots (0-chains) represent discrete informational states, whereas higher-dimensional cycles represent coherent transformations or flows among them.
The organizing law that makes these components compatible is the topological closure condition
$\partial^{2}=0$,
ensuring that local transitions assemble into globally valid computations without residual boundary.
This is the fundamental constraint that turns raw dynamics into a consistent inferential process.
Applying $\partial^{2}=0$ to a finite chain complex yields the Euler-Poincar\'{e} identity \citep{hatcher2005algebraic},
$\chi(\mathcal{K})
= \sum_{k=0}^{n} (-1)^k b_k
= \sum_m (b_{2m} - b_{2m+1})$,
where $b_k$ is the $k$-th Betti number.
The alternating sign structure expresses a bookkeeping law: even-indexed and odd-indexed Betti numbers enter $\chi$ with opposite sign and must balance for global closure.

\noindent\textbf{From Euler balance to functional dichotomy.}
The alternating sign in the Euler identity does not, by itself, assign distinct computational roles to even- and odd-dimensional homology; in standard algebraic topology, $b_2$ and $b_1$ are equally ``discrete'' invariants of a finite complex.
What \emph{does} establish a functional asymmetry is the special status of the lowest-dimensional pair, $b_0$ and $b_1$:
\begin{itemize}
    \item $b_0$ (connected components) counts the discrete, locally detectable structural units of the space - the tokens, basins, or attractors that anchor inference.
    These are stable under arbitrary continuous deformations within each component.
    \item $b_1$ (independent loops) counts the recurrent pathways - the cycles, oscillatory trajectories, and temporal flows that navigate among and within components.
    Each generator of $H_1$ supports a continuous family of representative cycles, giving $b_1$ a dynamical character absent from $b_0$.
\end{itemize}
The $b_0$/$b_1$ dichotomy is the \emph{dominant instance} of a broader functional partition between structural stability and dynamical flexibility.
We adopt the notation $\Heven$ (scaffold) and $\Hodd$ (flow) as shorthand reflecting this canonical pair, while noting that the extension to higher Betti numbers ($b_2, b_3, \ldots$) is a modeling hypothesis supported by the neuroscientific evidence below, not a theorem of algebraic topology.

\begin{table}[h]
    \centering
    \begin{center}
\small
\begin{tabular}{lcc}
\toprule
& \textbf{Scaffold} ($\Heven$) & \textbf{Flow} ($\Hodd$) \\
\midrule
Canonical generator & $\beta_0$ (components) & $\beta_1$ (cycles) \\
Neural code & Rate coding & Phase coding \\
Information type & Content / ``what'' & Context / ``where'' \\
Entropy regime & Low (discrete, persistent) & High (continuous, transient) \\
Temporal dynamics & Slow consolidation & Fast adaptation \\
\bottomrule
\end{tabular}
\end{center}
    \caption{Scaffold-flow model corresponds to rate/phase coding hypotheses in computational neuroscience.}
    \label{tab:parity}
\end{table}

\noindent

\noindent\textbf{Neuroscientific grounding: rate coding vs.\ phase coding.}
The scaffold-flow partition aligns with two well-characterized neural coding strategies \citep{dayan2005theoretical} (Table \ref{tab:parity}).
\emph{Rate coding} conveys information via the frequency of spikes integrated over a time window, discarding precise temporal structure \citep{van2001rate}.
Rate-coded representations are stable, slowly varying, and context-invariant - they encode \emph{what} is present.
\emph{Phase coding} conveys information via the precise timing of a spike relative to a background oscillation (e.g., hippocampal theta rhythm \citep{lisman2013theta}).
Phase-coded representations are transient, rapidly reconfigurable, and context-dependent, encoding \emph{where} or \emph{when} within an ongoing process.
Rate-coded ensembles define the discrete attractor landscape ($\beta_0$ structure) on which phase-coded trajectories ($\beta_1$ dynamics) unfold \citep{buzsaki2006rhythms}.
The theta-gamma oscillatory coupling observed in hippocampal circuits \citep{lisman2013theta} physically instantiates the scaffold-flow alternation: gamma bursts (fast, flow-like) are nested within theta cycles (slow, scaffold-like), and their interaction governs both encoding and retrieval. The parity-based correspondence is firmly supported by both mathematical necessity and computational implication.
A quantitative consequence of this nesting architecture is derived in
Sections \ref{sec:sfm} and~\ref{sec:mai}: the optimal condensation ratio between adjacent
oscillatory levels is $\rho = e$ (Corollary~\ref{cor:optimal_rho}), and the
resulting theta-gamma frequency ratio $e^2 \approx 7.4$ predicts the storage
capacity of working memory from first principles
(Proposition~\ref{prop:working_memory}).

\noindent\textbf{Connection to the Urysohn separator construction.}
The necessity of two-phase alternation is a structural necessity for analog computation \cite{mead_analog_2012}.
In parallel to the read/write logic in Turing machine \cite{sipser_introduction_1996}, the constructive proof of Urysohn's Lemma \citep{willard2012general} builds a separating function by iterating between two strokes: proposing open neighborhoods $U_r$ (\emph{expansion}) and enforcing closure containment $\overline{U_r} \subset U_s$ (\emph{contraction}).
The flow phase ($\Hodd$) implements the expansion stroke: the system explores, proposes candidate neighborhoods, and exposes boundary failures (points that should be separated but are not yet).
The scaffold phase ($\Heven$) implements the contraction stroke: the system takes closures, validates proposed neighborhoods, and consolidates them into compact tokens by contracting the metric within each validated region.
Each expansion-contraction cycle increases the margin $\delta$ between competing class supports; the alternation is the minimal mechanism that constructively manufactures the topological disjointness required for Urysohn's Lemma to activate.

\noindent\textbf{Computation as parity-regulated closure.}
Cognitive computation emerges as the enforcement of equilibrium between the content scaffold and the contextual flow.
Topologically, the closure law $\partial^{2}=0$ ensures that every contextual deformation ultimately reattaches to its supporting content scaffold without leaving a residual boundary \citep{rao1999predictive}.
This prevents dynamic context from destabilizing structural content, while content simultaneously constrains context to evolve along admissible recurrent pathways.
Informationally, the boundary condition can be viewed as a \emph{minimization of joint uncertainty} \citep{cover_elements_1999} between the structural and dynamical modes: a deformation that fails to close generates prediction error (boundary residual) or excess entropy.
The system descends an information gradient by driving $\partial c \to 0$ for all admissible deviations $c$, reducing the mismatch between what persists (content) and what varies (context).
Therefore, the parity principle identifies efficient cognition with the maintenance of a homological equilibrium: topological closure as the physical mechanism of biological memory formation and consolidation \citep{squire2015memory}, minimizing the mutual inconsistency between scaffold and flow, or equivalently, between content and context.

\subsection{Amortized Inference: From Dynamics to Structure}

If the parity principle describes the \emph{state} of satisfied cognition, how is this state achieved efficiently? A purely online inference system (like a standard Monte Carlo sampler \citep{liu2001monte}) must resolve the boundary equation $\partial c = 0$ from scratch for every new stimulus, which is computationally expensive and slow. This has been known as Minsky's search problem \citep{minsky1967computation} in AI or the curse of dimensionality \citep{bellman1966dynamic} in optimization. 
Inspired by Savitch's theorem \citep{savitch1970relationships} and Mountcastle's column organization principle \citep{mountcastle1957modality}, we introduce \textbf{amortized inference} \citep{gershman2014amortized} as the operational mechanism by which the brain converts transient dynamics into durable structural priors. The process of amortized inference follows a specific trajectory we term the \textbf{Topological Trinity Transformation}:  \textbf{Search $\to$ Closure $\to$ Condensation}.
1) \textbf{Search (Dynamics):} The network explores the energy landscape, attempting to resolve high-entropy contextual input ($\Psi$) and prediction errors (open boundaries).
2) \textbf{Closure (Topology):} The system identifies a recurrent path or attractor that neutralizes the error, forming a closed cycle ($\partial c = 0$) and establishing temporary coherence.
3) \textbf{Condensation (Plasticity):} Through synaptic consolidation such as Spike-Timing Dependent Plasticity (STDP) \citep{caporale2008spike} and episodic replay \citep{buzsaki1996hippocampo}, the transient cycle is embedded into the hard wiring of the network.

Specifically, we demonstrate that cognition operates by converting high-complexity recursive search (modeled by \textit{Savitch’s Theorem} \citep{savitch1970relationships} in NPSPACE) into low-complexity lookup (modeled by \textit{dynamic programming} \citep{bellman1966dynamic} in P) via the mechanism of \textbf{topological cycle closure}.
This amortization process parallels established biological and cognitive models. The alternating phases map naturally onto the \textbf{Wake-Sleep algorithm} \citep{hinton1995wake}, where wakeful exploration generates training data (cycles) that are consolidated into synaptic weights during sleep. Similarly, it provides a rigorous topological basis for the distinction between \textbf{Fast (System 1)} and \textbf{Slow (System 2)} thinking \citep{kahneman2011thinking}: System 1 represents the effortless execution of amortized invariants in $\Heven$, while System 2 corresponds to the energy-intensive search for parity in $\Hodd$.
Whereas naive dynamics search the energy landscape for a solution, established synaptic connectivity provides a topology of least resistance \citep{feynman2011feynman}. Through this process, context becomes content, and the cost of future inference is amortized \citep{gershman2014amortized}.


Taken together, the parity principle and amortized inference form a 
unified account of how the brain maintains coherence across space, time, and scale.  
The Parity Principle provides the structural foundation: even-dimensional homology encodes 
the \emph{content scaffold}, the stable, low-entropy structures that anchor memory \citep{tulving1972episodic}, 
while odd-dimensional homology encodes the \emph{contextual flow}, the dynamic, 
high-entropy modulations that adapt those structures to new conditions.  
This scaffold-flow decomposition can be applied recursively to create a ``Tower of Scaffolds'' and becomes the topological substrate on which \textbf{memory-amortized inference} (MAI) operates.
MAI gives the dynamical realization: inference proceeds by alternating between 
retrieval onto the scaffold and bootstrapping along contextual flows.  
Such alternation serializes nondeterministic exploration into deterministic cycles, 
ensuring that every deformation ultimately reattaches to its structural support.  

In this view, biological intelligence is not a mechanism for computing arbitrary 
states, but a mechanism for \emph{closing loops} \citep{varela2025principles}: enforcing $\partial^{2}=0$ across 
a multiscale hierarchy of interacting cycles.  
The resulting homological backbone enables memory reuse, stabilizes perceptual 
interpretation, and makes fast inference possible on a slow physical substrate \citep{anderson2013architecture}.
The two principles integrate naturally:  
parity provides the \emph{architecture} (scaffold vs.\ flow),  
while amortization provides the \emph{algorithm} (alternation vs.\ recurrence).  
Parity guarantees that content and context remain mutually consistent, and MAI supplies the dynamical routine for reducing their mismatch over time.  
Together, they predict that violations of this equilibrium, failures to close loops, will manifest as characteristic cognitive impairments \citep{petersen2001current}.  
Insufficient amortization (weak consolidation of loops) leads to unstable or rapidly 
shifting percepts, as seen in schizophrenia \citep{insel2010rethinking}, whereas excessive rigidity (overstabilized 
scaffolds) produces inflexible priors and reduced contextual modulation, as in autism \citep{baron1985does}.  
These conditions represent opposite failures of the same underlying law: the failure to 
maintain parity-regulated closure between scaffold and flow.
The MAI framework reframes the brain not as a device that \emph{computes states},  
but as one that \emph{stabilizes cycles}.  
By using topological structure to transform local interactions into global 
understanding, the brain achieves what would otherwise be computationally impossible:  
performing rich inference through the persistent enforcement of closure between 
what persists (scaffold) and what varies (flow).

\section{Scaffold-Flow Memory Model via the Parity Principle}
\label{sec:sfm}


\subsection{Neural States as Topological Chains}
\label{subsec:parity}
We ground our topological model in the biophysics of \textbf{Polychronous Neural Groups (PNGs)} \citep{izhikevich2006polychronization}. PNGs represent reproducible, time-locked sequences of spikes that emerge in recurrent spiking networks with fixed axonal delays and spike-timing-dependent plasticity (STDP) \citep{caporale2008spike}. These groups form the functional building blocks of neural computation, allowing the network to encode relationships across time as spatial patterns of connectivity.
At the theoretical foundation, we introduce the notion of a \emph{spatiotemporal complex} \citep{edelsbrunner2022computational}, which captures temporally consistent neural interactions within a finite window, and relate this structure to a classical \emph{chain complex} in algebraic topology \citep{hatcher2005algebraic}, allowing us to extract persistent topological features corresponding to functional PNGs.
Let $S = \{n_1, n_2, \dots, n_N\}$ be a set of neurons. A $k$-simplex $\sigma_k = [n_0, \dots, n_k]$ represents a functional co-activation (a PNG) binding $k+1$ neurons into a coherent signaling unit. To capture the dual nature of neural information processing, we generalize the state space from scalar activations to complex phasors. The instantaneous state of the neural system is described by a complex chain $c_k$, a formal sum of active simplices: $c_k = \sum_{j} z_j \sigma_k^j, \text{where} z_j = \rho_j e^{i \theta_j}$, where the complex coefficient $z_j$ encodes the parity-alternating logic of the neural code: 1) The modulus ($\rho_j \in \mathbb{R}^+$), representing the firing rate (Intensity), acts as the $\Heven$ Scaffold component, quantifying the structural \textit{existence} or magnitude of the feature (Integration); 2) The Phase ($\theta_j \in S^1$), representing the spike timing, acts as the $\Hodd$ flow component, encoding the temporal \textit{relation} or trajectory of the feature (Resonance). Our formulation ensures that every topological feature carries both a static weight ($\rho$) and a dynamic vector ($\theta$), allowing the system to perform spatial integration and temporal resonance simultaneously via complex chain operations.

Each PNG is defined by a set of neurons whose spikes
occur in precise temporal relations, forming a reproducible spatiotemporal
pattern.  Such patterns naturally induce a \emph{cell poset}: every neural event is a
cell, and the causal-temporal relation ``$e_i$ precedes $e_j$'' defines an order
relation $e_i \preceq e_j$.  The resulting poset encodes both the geometry
(who is connected) and the dynamics (who influences whom) of the PNG.  
By passing to the order complex $\Delta(P)$, this poset becomes a simplicial
complex, and the neural pattern becomes a \emph{topological chain}: a sum of cells
whose boundaries represent the flow of neural influence.  
This mapping provides a universal representational format for neural dynamics:
any temporally extended firing pattern can be expressed as a chain in a cell poset.
By passing to the order complex $\Delta(P)$, this poset becomes a simplicial
complex, and the neural pattern becomes a \emph{topological chain}: a sum of cells
whose boundaries represent the flow of neural influence.  
This mapping provides a universal representational format for neural dynamics \citep{gerstner2014neuronal}:
any temporally extended firing pattern can be expressed as a chain in a cell poset.
\emph{Geometric realization of these simplicial structures leads to a continuous
latent space representation}, in which recurrent neural patterns tile a manifold
$\mathcal{Z}$ whose associated chain complex $(C_\bullet(\mathcal{Z}),\partial)$
is the continuum limit of the discrete cell complex \citep{spanier1966algebraic}.
Local closure ($\partial^2=0$)
ensures consistency (integration), while persistent homology quantifies the
diversity of coexisting invariants (differentiation).
In this framework, $|\chi|$ functions as a form of
\emph{topological free energy} \citep{friston2010free} that is minimized during inference to enforce
global consistency, whereas $\mathcal{I}_{\mathrm{global}}$ serves as a
measure of \emph{topological evidence} accumulated across development and
learning.  


\subsection{The Scaffold-Flow Model of Memory}
\label{subsec:scaffold_flow}

Projecting the parity principle onto biological architecture yields the scaffold-flow model. Memory is not a static store but a homological filtration process defined by the decomposition of a memory trace $\gamma_i$:
\begin{equation}
\label{eq:homological_memory_model}
\gamma_i
= \underbrace{\sigma}_{\text{Scaffold } (\Phi)}
  \;+\;
    \underbrace{\sum_{k=1}^{b_i} a_{ik}\,\beta_k}_{\text{Flow } (\Psi)}
  \;+\;
    \underbrace{\partial d_i}_{\text{Noise}}
\end{equation}

\begin{itemize}\item \textbf{Scaffold ($\Phi=\sigma$):} The stable backbone ($\partial \sigma=0$) common to multiple traces. It acts as a low-pass filter, preserving invariant structure (semantic memory \citep{binder2011neurobiology}).\item \textbf{Flow ($\Psi$):} A linear combination of basis loops $\{\beta_k\}$ representing specific exploratory trajectories (episodic memory \citep{tulving1972episodic}).\item \textbf{Noise ($\partial d_i$):} The boundary term. By the law $\partial^2=0$, the noise component is topologically trivial and transient, representing sensory details discarded during consolidation.\end{itemize}

\noindent\textbf{Memory Consolidation as Topological Condensation.} The scaffold-flow model reinterprets memory consolidation not as data transfer, but as a topological phase transition implemented by generalized Hebbian learning. When a novel event (a high-fidelity $\Hodd$ cycle) is consolidated \citep{squire2015memory}, the system extracts its invariant structure. The specific temporal links are relaxed, and the persistent features are frozen into the $\Heven$ scaffold (e.g., $H_1$-cycle becomes a $H_0$-dot after condensation). With external sensory noise absent, $\Hodd$ content, dominated by replayed episodic memory traces \citep{wilson1994reactivation}, provides training data to anneal the $\Heven$ scaffold. 
Short-term traces, dominated by transient chains ($\partial d_i$) within the spatiotemporal complex $\mathcal{K}_\delta(t)$, correspond to high-entropy sensory noise. Through recurrent replay, these transient boundaries are annihilated ($\partial d_i \to 0$) via the cycle-closure principle, leaving behind only the persistent chains ($\sigma + \sum a_{ik}\beta_k$) that form the stable memory manifold. 

Through recurrent replay and synchronization \citep{lisman2013theta}, these short-term traces undergo topological averaging, progressively stripping away non-recurrent boundaries ($\partial d_i \to 0$) to reveal the invariant loops that satisfy $ \partial^2 = 0 $. The consolidation process refines the $\mathcal{H}_{\text{even}}$ scaffold ($\sigma$) through homological promotion. If a specific contextual loop $\beta_k$ persists across multiple disjoint episodes, the system condenses it into the backbone itself ($\sigma_{new} \leftarrow \sigma_{old} \cup \text{Condense}(\beta_k)$). Under the structure-before-specificity (SbS) framework, consolidation amounts to a dual operation: 1) filtration - converting transient boundaries ($ \gamma \in \operatorname{im}\partial_2 $) into persistent homology classes ($ [\gamma] \in H_1(\mathcal{K}_\delta) $) \citep{edelsbrunner2008persistent}; 2) deformation - embedding high-entropy contextual flows ($ \Psi=\Hodd$) that survive filtration into the low-entropy structural scaffold ($ \Phi=\Heven $). The resulting long-term memory (LTM) trace \citep{shiffrin1969storage} is not a static replica of the original event but a structural attractor, retrievable through the completion of a homological loop rather than the reactivation of an exact firing pattern.

\subsection{Biological Signatures: PNGs and Columns}
\label{sec:png_columns}

The scaffold–flow decomposition provides a static description of neural memory
structures, but cognition is fundamentally a \emph{dynamic} process.  The brain
does not merely store $\sigma$ and $\beta_k$; it continuously \emph{moves
between} them.  What makes the scaffold–flow model powerful is that it
naturally predicts two complementary modes of operation, depending on the
direction of information flow between even-dimensional scaffolds
($\Heven$) and odd-dimensional flows ($\Hodd$).  When flow collapses into
scaffold, the system stabilizes structure and consolidates knowledge; when
scaffold expands into flow, the system generates predictions and explores
interpretations.  These two modes, one compressive and the other generative, emerge
directly from the homological architecture encoded in
Eq.~\eqref{eq:homological_memory_model}.  We now describe how these dual
operations correspond to the biological phases of \emph{learning} and
\emph{inference}, as shown in Fig. \ref{fig:two_phase_condensation_expansion}.

\noindent\textbf{Learning as Topological Condensation (Sleep).}
Learning constitutes a global, structural phase transition in which dynamic
\textit{Contextual Flow} ($\Psi \in \Hodd$) is progressively absorbed into a
stable \textit{Content Scaffold} ($\Phi \in \Heven$). Novel experiences initially
manifest as high-entropy, odd-dimensional trajectories, $\beta_1$-type flows
capturing specific episodic sequences. Across repeated activation or offline
replay, these transient loops undergo a ``freezing'' operation: their temporal
dimension is integrated out, collapsing $\beta_1$ trajectories into
lower-dimensional scaffold components. Formally, each replayed trace $\gamma_i$
is decomposed according to Eq.~\eqref{eq:homological_memory_model}, and the
residual boundary $\partial d_i$ quantifies its failure to close against the
current scaffold. Sleep-dependent learning (e.g., sharp-wave ripples (SWRs) \cite{wilson1994reactivation}) minimizes these residuals by updating
both the coefficients $a_{ik}$ and the scaffold $\sigma$ so that
$\partial d_i \to 0$ across traces. 
This Structure-before-Specificity (SbS)
process anneals the scaffold into a lower-entropy, more coherent manifold that
encodes the shared structure distilled from episodic variability. Therefore,
Topological condensation accumulates negentropy by transforming the metabolic cost of
time and motion \cite{landauer1961irreversibility} into a reusable structure, embedding contextual flows into a persistent semantic backbone.
In summary, we propose that the abstract parity distinction between Flow and Scaffold maps directly onto the known physiological distinction between temporal firing patterns and anatomical modules.

\noindent\textbf{Odd and Even Parity: PNGs and Cortical Columns.}
The biological signature of the $\Hodd$ regime is the PNG\citep{izhikevich2006polychronization}. PNGs are time-locked spatiotemporal patterns where neurons fire in precise sequences determined by axonal conduction delays.
Topologically, a PNG is a \textbf{1-cycle} ($\beta_1$): it is a causal chain of events that must be traversed in time. It represents the active \emph{search path} or inference trajectory. Just as $\Hodd$ flows are high-entropy and context-sensitive, PNGs are capable of encoding vast combinatorial possibilities ($\text{NPSPACE}$) but are transient and metabolically expensive to sustain.
The biological signature of the $\Heven$ regime is the \textbf{Cortical Column} \citep{mountcastle1997columnar}. If PNGs represent the ``software'' of dynamic flow, the column represents the ``hardware'' of static storage.
Topologically, the column acts as the \textbf{0-cycle} ($\beta_0$): a spatial vertex that integrates inputs. Through the process of condensation (sleep mode), a stable, recurring PNG (a closed temporal loop) is physically collapsed via STDP \citep{caporale2008spike} into the strong synaptic weights of a specific columnar assembly, which confirms the trinity transformation in wetware:
$\underbrace{\text{PNG}}_{\text{Time/Flow } (\beta_1)} \xrightarrow{\text{STDP}} \underbrace{\text{Columnar Assembly}}_{\text{Space/Structure } (\beta_0)}$.
The brain essentially uses PNGs to search for causal relationships in time, and once found, freezes them into the spatial architecture of columns.
Inference constitutes the complementary, dynamical phase in which the scaffold is ``melted'' back into flow. Here, the stable \textit{Content Scaffold} ($\Phi \in \Heven$) is projected into dynamic \textit{Contextual Flow} ($\Psi \in \Hodd$) to navigate the environment. A low-entropy invariant, such as a concept encoded in a $\beta_0$ component, is ``unpacked'' into a high-dimensional $\beta_1$ trajectory that simulates expected sensations. Crucially, this process operates as a rapid \emph{topological pulse}: first, \textbf{expansion}, where the scaffold generates a cloud of admissible hypotheses (branching search); followed by \textbf{contraction}, where bottom-up sensory constraints force the flow to snap onto a single, self-consistent cycle ($\partial c=0$). This scaffold-flow alternation implements a \emph{context-before-content} loop: the scaffold restricts exploration to low-energy manifolds, while the flow updates the scaffold's predictions by selecting the most coherent trajectory. 
\begin{figure}[t]
\centering
\begin{tikzpicture}[
    >=Latex,
    font=\small,
    phasebox/.style={rounded corners, draw, thick, fill=black!4,
                     minimum width=4.0cm, minimum height=0.9cm, align=center},
    cloud/.style={ellipse, draw, thick, fill=black!2,
                  minimum width=3.2cm, minimum height=1.1cm, align=center},
    scaffold/.style={circle, draw, thick, fill=black!2, minimum size=1.0cm},
    lab/.style={font=\scriptsize, align=center}
]

\node[phasebox] (sleep) at (0,2.4)
  {Sleep / Learning\\Topological Condensation};

\node[cloud] (flowsL) at (0,1.0)
  {Contextual flows\\$\gamma_i,\,\beta_k \in H_{\mathrm{odd}}$};

\node[scaffold] (scafL) at (0,-1.) {$\sigma$};
\node[lab,below=0.1cm of scafL]
  {Content scaffold\\$\Phi \in H_{\mathrm{even}}$};

\draw[->,thick] (flowsL) -- node[lab,right,xshift=1pt]
  {condensation\\$\partial d_i \to 0$} (scafL);

\node[lab,left=0.2cm of flowsL.west]
  {replay\\(offline)};

\node[phasebox] (wake) at (7.0,2.4)
  {Wake / Inference\\Topological Expansion};

\node[scaffold] (scafR) at (7.0,-1.) {$\sigma$};
\node[lab,below=0.1cm of scafR]
  {Content scaffold\\$\Phi \in H_{\mathrm{even}}$};

\node[cloud] (flowsR) at (7.0,1.0)
  {Contextual flows\\$\Psi,\,\beta_k \in H_{\mathrm{odd}}$};

\draw[->,thick] (scafR) -- node[lab,right,xshift=1pt]
  {expansion\\prediction} (flowsR);

\node[lab,left=0.2cm of flowsR.west]
  {simulation\\(online)};

\draw[<->,thick] (sleep.east) -- node[lab,above]
  {scaffold-flow cycle} (wake.west);

\node[lab,below=-0.5cm of wake.south east,anchor=north east,xshift=-4.5cm]
  {$\partial^2 = 0$};

\end{tikzpicture}
\caption{Two-phase scaffold-flow dynamics. 
Left: during sleep, topological condensation transforms episodic contextual flows 
(replay trajectories in $H_{\mathrm{odd}}$) into refinements of the content scaffold 
$\sigma \in H_{\mathrm{even}}$ by driving residual boundaries $\partial d_i \to 0$. 
Right: during waking inference, topological expansion “melts’’ the scaffold back 
into contextual flows, using stored structure to generate predictions and explore 
admissible interpretations. Together, these phases form a bidirectional cycle that 
enforces multiscale closure $\partial^2 = 0$.}
\label{fig:two_phase_condensation_expansion}
\end{figure}


\subsection{Recursive Condensation: Building the Tower of Scaffolds}
\label{subsec:tower_scaffold}

\noindent\textbf{From Flow to Scaffold}
In standard neural network theory, layers are often viewed as performing identical filtering operations at different scales. In our framework, the relationship between layers is transformational.
The output of layer $L_k$ is not a feature map but a topological phase transition.
\begin{itemize}
    \item At Level $k$ (Flow / $\Psi$): A pattern exists as a temporal correlation - i.e., a rapid sequence of firing (a cycle $\beta_1$). The system must expend energy to maintain this relationship.
    \item At Level $k+1$ (Scaffold / $\Phi$): That same pattern exists as a spatial address, such as a single neuron or microcolumn (a vertex $\beta_0$). The system treats it as a static atom.
\end{itemize}

We formalize a level-to-level transition as the conversion of odd-dimensional homology (cycles) into even-dimensional homology (components).

\begin{proposition}[Recursive Condensation]
\label{prop:recursion}
The topological structure of the cortical hierarchy is generated by the recursive transformation of homological flow into a structural scaffold:
$\mathcal{H}_{\text{odd}}^{(k)} \xrightarrow{\text{Condense}} \mathcal{H}_{\text{even}}^{(k+1)}$
Specifically, a stable, high-frequency limit cycle ($\beta_1$) at level $k$ is treated as a static atomic unit ($\beta_0$) at level $k+1$.
\end{proposition}


\begin{corollary}[Optimal Condensation Ratio]
\label{cor:optimal_rho}
Among all continuous hierarchical manifolds satisfying
Proposition~\ref{prop:recursion}, the total inference action is minimized
when the condensation ratio $\rho$ between adjacent Tower levels equals
Euler's number $e$.
\end{corollary}

\begin{proof}
Let $N$ denote the state-space complexity of the environment and let
$\rho > 1$ be the condensation ratio: the number of $\Hodd$ micro-states at
level $k$ that must chain together before condensing into a single $\Heven$
macro-state at level $k+1$.  By Proposition~\ref{prop:recursion}, the Tower
depth spanning $N$ states is $D(\rho) = \log_\rho N = \ln N / \ln \rho$.
Each level imposes an intra-level metric cost proportional to the basin size
$\rho$ (the number of Urysohn boundary evaluations required to verify
membership in one equivalence class at that level; see the Urysohn separator
construction in Section~\ref{subsec:parity}).  The total inference action is
therefore
$\mathcal{C}(\rho)
    = \rho \cdot D(\rho)
    = \ln N \cdot \frac{\rho}{\ln \rho}$.
Differentiating with respect to $\rho$ and setting
$d\mathcal{C}/d\rho = 0$:
$\frac{d\mathcal{C}}{d\rho}
    = \ln N \cdot \frac{\ln\rho - 1}{(\ln\rho)^2} = 0
    \;\implies\; \ln\rho = 1
    \;\implies\; \rho = e$.
The second derivative $\mathcal{C}''(e) = e^{-1} > 0$ confirms a strict
global minimum on $\rho > 1$.  At the optimum,
$\mathcal{C}(e) = e\ln N$ is the minimum achievable inference action for a
continuous hierarchical manifold of complexity $N$.
\end{proof}

\begin{remark}[Biological instantiation and robustness]
\citet{penttonen2003natural} measured the center frequencies of mammalian
cortical oscillatory bands and found that adjacent bands maintain a nearly
constant multiplicative ratio of approximately $e \approx 2.718$, spanning
more than four orders of magnitude from ultra-slow rhythms (${\sim}0.05$\,Hz)
to ultra-fast ripples (${\sim}600$\,Hz) \citep{buzsaki2006rhythms}.
Corollary~\ref{cor:optimal_rho} provides a first-principle thermodynamic
explanation: the $e$-ratio is not an arbitrary biological constant but the
unique optimal condensation ratio for a continuous hierarchical inference
manifold.  The cost function $\mathcal{C}(\rho)$ is nearly flat on the
interval $[e^{0.9},\,e^{1.1}] \approx [2.46,\,3.00]$, incurring less than
$5\%$ excess action relative to the minimum; biological systems can therefore
deviate slightly from $e$ without significant efficiency loss, accounting for
the observed inter-band ratio variance in electrophysiological recordings.
In the discrete limit, the nearest integer to $e$ is $3$, recovering the
known result from information theory that base-3 is the most efficient
integer radix \citep{hayes2001computing}; the brain, operating as a
continuous analog system, achieves the exact theoretical optimum.
\end{remark}

\noindent\textbf{The Tower of Scaffolds.}
The recursion introduced above resolves a classical difficulty for any theory of high-level cognition: Minsky's search problem \citep{minsky1967computation}. Without hierarchical reuse, representing a concept like ``face'' would require co-activating every human interaction that defines it - a combinatorial load no biological substrate can sustain in real time.

Recursive condensation makes that load bounded. The brain builds a literal tower of scaffolds, most explicitly traced in the ventral visual stream \citep{dicarlo2012does}:
\begin{enumerate}
    \item V1 (pixels $\to$ edges): correlations of light ($\beta_1$) condense into line segments ($\beta_0$).
    \item V4 (edges $\to$ shapes): closed cycles of lines ($\beta_1$) condense into geometric primitives ($\beta_0$).
    \item IT (shapes $\to$ objects): configurations of shapes ($\beta_1$) condense into semantic entities ($\beta_0$).
    \item PFC (objects $\to$ concepts): causal loops between entities ($\beta_1$) condense into abstract rules ($\beta_0$).
\end{enumerate}
By the time the signal reaches the prefrontal cortex \citep{fuster2008prefrontal}, the atoms of computation are no longer sensory details but entire causal histories compressed into single points. Each level uses the same local condensation dynamics as the level below; what changes is the identity of what counts as an atom. The agent can then manipulate complex narratives with the same low-latency efficiency with which early visual cortex manipulates edges, the operational basis for zero-shot generalization \citep{brown2020language}.

The same tower architecture explains why depth matters in deep learning \citep{ayzenberg2025sheaf}. A shallow network such as a multi-layer perceptron \citep{rosenblatt1958perceptron} attempts to map input to output in a single condensation step; if the causal depth of the problem exceeds the depth of the network, the required ``wormhole'' is too long, and error compounds along it. A deep network breaks the long geodesic into short verifiable hops ($\beta_1 \to \beta_0 \to \beta_1 \to \beta_0$), building stable intermediate scaffolds at every level. Depth, on this reading, is not a parameter-count heuristic but a \emph{topological resolution}: the number of parity alternations the substrate can sustain before closure fails. Intelligence is the height of the tower one can maintain.
This raises a sharper question. Each additional level of the tower costs only polynomial hardware, one more cortical column, one more transformer block, yet the representational reach grows exponentially, because the set of states expressible at level $k+1$ branches combinatorially over the already-condensed states of level $k$. What formal principle licenses this polynomial-in-height, exponential-in-reach architecture, and what does it demand of the brain's memory substrate? The next section argues that the answer is a space-time tradeoff of the form identified by Savitch's theorem, made biological through memory amortization.

\section{Trading Space with Time: Amortized Inference}
\label{sec:mai}


\noindent\textbf{The Paradox.}
Savitch’s theorem $NSPACE(f(n))\subseteq DSPACE(f(n)^2)$ showed that nondeterministic branching can be \emph{serialized}
without exponential space cost \citep{savitch1970relationships}.
But the combinatorial explosion “what-if’’ simulations persist, which renders high-level cognition intractable under $\mathrm{P}\neq\mathrm{NP}$.
Vernon Mountcastle’s discovery in the 1950s \citep{mountcastle1957modality} found that the neocortex is built from a single repeating microcircuit, the
\emph{canonical cortical column}, whose architecture is replicated across all
sensory and associative areas.
Each new column adds only a small amount of biological hardware (neurons and
synapses), but the overall network gains access to exponentially more composite
states and relational structures.
This empirical fact biologically implements Savitch's principle of \emph{structural
amortization}: polynomial growth of substrate yields exponential growth of
representational capacity.
If each cortical column implements roughly the same local dynamics,
why does adding more columns, say, doubling the cortical area, yield
orders of magnitude increase in cognitive power?

\noindent\textbf{The Solution.}
Each column contributes not a new function, but a new \emph{dimension of
coupling}. As these modules interconnect, the number of possible
topologically distinct PNGs grows combinatorially with the number of intercolumnar links.
A small increase in column count $N$ produces
$b_1 \sim O(N^2)$,
depending on the degree of recurrent interconnectivity $\alpha$.
A linear expansion of structural resource can support exponential
growth $O(e^{\alpha N})$ in representational loops, the latent search space for cognition.
The brain defeats the exponential complexity of $\mathrm{NP}$-hard inference by transforming the problem through our scaffold-flow memory model. Specifically, we draw a structural analogy between Savitch's recursive serialization of nondeterministic space \citep{savitch1970relationships} and the reuse of condensed cycles across inference episodes \citep{mountcastle1957modality}. 
Instead of explicitly enumerating all counterfactual branches, the neocortex 
\emph{encodes the structure of the search space} as a distributed recurrent manifold of stable cycles.  
Each cycle $\gamma_i$ represents a locally consistent hypothesis, while their common intersection preserves global coherence across contexts.  
The cortex represents multiple ``what-if'' trajectories \cite{minsky1967computation} as concurrent deformations around the shared structural scaffold. Amortized inference \citep{gershman2014amortized} makes this representation
computationally effective. The alternation of $\mathcal{F}$ (bootstrapping) and
$\mathcal{R}$ (retrieval) implements a deterministic simulation in which each
hypothesis is explored by deforming the current state along an admissible loop
$b_k$ and then re-projecting onto the backbone $\sigma$. The recursive
departure-and-return pattern along the scaffold-flow model is \emph{structurally analogous} to Savitch's serialization:
a nondeterministic branching search is replaced by a deterministic procedure that reuses a
bounded amount of structural memory while preserving boundary relations
($\partial^2=0$). 

\subsection{Memory-Amortized Inference (MAI): A Unified Framework}

Based on the memory model and amortized inference, we now introduce MAI as the operational logic of the homological brain. MAI is defined as the process of minimizing the topological free energy $|\chi|$ by converting high-complexity search into low-complexity storage.
We formalize MAI as a general strategy for reducing the computational cost of inference by storing and reusing structured latent representations. The key idea is to construct a memory of prior inference results such that new inference problems can be approximated by querying and adapting from this memory, rather than solving the full problem from scratch.
Let \( \Psi \in \mathcal{X} \) denote the observable context and \( \Phi \in \mathcal{S} \) the latent content to be inferred. Let \( \mathcal{L}(\Psi, \Phi) \) denote a loss or cost function encoding the fidelity or predictive value of \( \Phi \) under context \( \Psi \). We assume that inference corresponds to solving the following optimization:
$\Phi^* = \arg\min_{\Phi \in \mathcal{S}} \left[ \mathcal{L}(\Psi, \Phi) \right]$.
Formally, we start with the following definition.

\begin{definition}[Memory-Amortized Inference]
Let \( \mathcal{M} = \{ (\Psi^{(i)}, \Phi^{(i)}) \}_{i=1}^N \) be a memory of prior context–content pairs, and let \( \mathcal{R}: \mathcal{X} \times \mathcal{M} \to \mathcal{S} \) be a retrieval-and-adaptation operator and $\mathcal{F}: \mathcal{S}\times\mathcal{X}\to\mathcal{S}$ be the bootstrapping update operator implemented via generative simulation. Inference is said to be \emph{memory-amortized} if it is formulated as a structural cycle between \emph{content} \( \Phi \) and \emph{context} \( \Psi \), where memory acts as a reusable substrate for inference:
$\Phi_{t+1} = \mathcal{F}(\Phi_t, \Psi_t), \quad \Phi_t \approx \mathcal{R}(\Phi_{t+1}, \Psi_t)$
in lieu of directly optimizing \( \Phi^* \), such that the expected cost satisfies
$\mathbb{E}_{\Psi} \left[ \mathcal{L}(\Psi, \hat{\Phi}) \right] \leq \mathbb{E}_{\Psi} \left[ \mathcal{L}(\Psi, \Phi^*) \right] + \varepsilon$,
for some amortization gap \( \varepsilon \ll \mathcal{L}(\Psi, \cdot) \), and where the runtime cost of \( \mathcal{R} \) is substantially lower than full inference.
\end{definition}

\noindent\textbf{The Retrieval-and-Adaptation Operator \(\mathcal{R}\).}
The retrieval-and-adaptation operator \( \mathcal{R}: \mathcal{X} \times \mathcal{M} \to \mathcal{S} \) serves as the core mechanism by which inference avoids re-computation. Given an input query (typically latent or perceptual), \( \mathcal{R} \) retrieves relevant elements from the memory \( \mathcal{M} = \{ (\Psi^{(i)}, \Phi^{(i)}) \}_{i=1}^N \) and performs a lightweight adaptation to generate a candidate solution \( \hat{\Phi} \). Operationally, \( \mathcal{R} \) consists of two stages:
1) \textbf{Retrieval:} Identify a relevant subset of memory entries \( \{ (\Psi^{(j)}, \Phi^{(j)}) \} \subset \mathcal{M} \) based on similarity to the current context \( \Psi_t \). This can be performed via kernel-based attention, similarity search in latent space, or topological proximity under homological constraints.
2) \textbf{Adaptation:} Modulate or interpolate the retrieved \( \Phi^{(j)} \) values conditioned on \( \Psi_t \), resulting in a candidate \( \hat{\Phi}_t = \mathcal{R}(\Phi_{t+1}, \Psi_t) \). This step often involves gradient-free adjustments (e.g., feature warping, parameter blending) and is significantly cheaper than full inference.

\noindent\textbf{The Bootstrapping Update Operator \(\mathcal{F}\).}
The bootstrapping operator \( \mathcal{F}: \mathcal{S} \times \mathcal{C} \to \mathcal{S} \) governs the internal dynamics of inference by iteratively updating the latent content representation \( \Phi_t \) given the context \( \Psi_t \). It defines a recurrence $\Phi_{t+1} = \mathcal{F}(\Phi_t, \Psi_t)$, where \( \mathcal{F} \) encodes the system’s structural prior, capturing the directionality, topology, and dynamic consistency of inference over time. Unlike standard update rules that minimize a loss from scratch, \( \mathcal{F} \) performs bootstrapping: each update is initialized from a prior memory-induced state. 
The bootstrapping update operator \( \mathcal{F} \) in MAI is structurally analogous to the \emph{half-step down} trick used in Q-learning \citep{watkins1992q} and temporal difference methods \cite{sutton2018reinforcement}, but in reverse time. While Q-learning bootstraps value via reward-driven transitions ($Q(s_t) \approx Q(s_{t+1})$), MAI bootstraps inference through latent memory and context ($\Phi_{t+1} = \mathcal{F}(\Phi_t, \Psi_t)$ inverted by $\Phi_t \approx \mathcal{R}(\Phi_{t+1}, \Psi_t)$). This dual relationship forms the backbone of the MAI half-step trick: the current latent content generates the next-step prediction, which in turn reconstructs the past, yielding a cycle-consistent structure that reduces entropy.

\noindent\textbf{MAI Paradigm: Search Builds Memory; Navigation Uses It.}
Within the MAI framework, slow,
resource-intensive \emph{search} corresponds to open-chain exploration
($\partial c \neq 0$) in the odd-dimensional flow space, used only when the
system lacks a suitable scaffold. Once a trajectory is closed and condensed
into an even-dimensional invariant, the agent no longer performs search but
\emph{navigates}. Inference becomes the fast traversal of a learned scaffold rather
than the exploration of a combinatorial tree. 
The dual operators of MAI provide a mechanistic account of the behavioral
dichotomy between slow and fast thinking \citep{kahneman2011thinking}.
During the retrieval-driven search phase, inference behaves like Savitch
recursion: the system explores counterfactual paths through a large latent
state space by invoking serial, resource-intensive simulations. This mode
corresponds to deliberate, slow cognition (System~2), whose hallmark is the
ability to navigate complex, branching contingency structures despite strict
capacity limits \citep{miller1956magical}. Once a trajectory has been closed and condensed,
the bootstrapping operator $\mathcal{F}$ implements dynamic programming (DP) in the
topological domain: previously costly simulations are replaced with direct
structural traversal of the scaffold, yielding rapid, automatic responses
(System~1). Therefore, MAI unifies slow and fast thinking as two operating regimes of
a single topological architecture: Savitch-style recursive search for novel
contexts, and DP-style amortized reuse for familiar ones.


\subsection{Topological Trinity Transform: From Savitch to Dynamic Programming}
We formalize MAI as a three-step topological transformation 
$\text{Search} \;\longrightarrow\; \text{Closure} \;\longrightarrow\; \text{Condensation}$,
which maps the temporal complexity of exploratory inference onto the spatial
efficiency of structural memory.  These phases directly correspond to the MAI
operators: \emph{Search} invokes the retrieval operator $\mathcal{R}$ to locate
a provisional scaffold against which the system tests hypotheses, while
\emph{Condensation} implements the bootstrapping operator $\mathcal{F}$ to
integrate validated flows back into the scaffold.

\noindent 1) \textbf{Search (Retrieval Phase).}  
When the system encounters a novel or ambiguous context $\Psi$, it first calls
the retrieval operator $\mathcal{R}$ to anchor inference onto an existing
portion of the scaffold. Because no fully compatible structure exists, the
system must engage in recursive simulation, an $\mathrm{NPSPACE}$-like process
consistent with Savitch’s theorem \citep{savitch1970relationships}.  
Formally,
$\Psi_{\mathrm{search}} \;=\; \mathcal{R}(\Psi,\Phi) \;\leadsto\; \text{Recurse}(\Hodd)$.
This generates a sequence of exploratory, high-entropy flows ($\partial c \neq
0$). Biologically, this corresponds to slow, deliberate inference (System 2),
where the system tests context against the scaffold and traces candidate paths
through the latent manifold.

\noindent 2) \textbf{Closure (Bootstrapping Phase).}  
Search terminates when a candidate trajectory becomes self-consistent with the
scaffold and closes into a cycle:
$\partial(\Psi_{\mathrm{trace}}) = 0$.
This event constitutes a topological phase transition: a formerly open,
uncertain trajectory becomes a persistent recurrent structure. Such closure is
the computational signature of insight \citep{bowden2005new}, marking the
transition from entropy to invariance. Only closed trajectories qualify for
integration into the scaffold.

\noindent 3) \textbf{Condensation (Amortization Phase).}  
Once a cycle is validated, the system applies the bootstrapping operator
$\mathcal{F}$ to incorporate this new recurrent pattern into the structural
scaffold. This operation performs \emph{topological condensation}: transforming
a high-dimensional, time-extended flow (an $\Hodd$ trajectory) into a
low-dimensional, stable, even-homology component ($\Heven$).  
Formally,
$\Phi_{\mathrm{new}}
\;=\;
\mathcal{F}(\Phi, \Psi_{\mathrm{closed}})
\;=\;
\text{Condense}(\Psi_{\mathrm{closed}})$,
which is mathematically equivalent to memoization in dynamic programming
\citep{bellman1966dynamic}. This step ``burns in'' the successful cycle,
replacing expensive recursive search with direct structural reuse. When the
context reappears, inference proceeds not by searching but by traversing the
updated scaffold, realizing amortized inference \citep{gershman2014amortized}.

\begin{proposition}[Theta-Gamma Nesting Capacity and Working Memory]
\label{prop:working_memory}
Let the cortical oscillatory hierarchy satisfy the optimal condensation ratio
$\rho = e$ of Corollary~\ref{cor:optimal_rho}.  Denote by
$\mathcal{L}_\theta$ and $\mathcal{L}_\gamma$ the Tower levels corresponding
to theta ($4$--$10$\,Hz) and gamma ($30$--$80$\,Hz) oscillations
respectively, with one intermediate level $\mathcal{L}_\beta$ (beta,
$10$--$30$\,Hz).  The number of gamma cycles $\mathcal{N}$ that fit within
one theta cycle, equivalently, the number of $\Hodd$ flow units that a
single $\Heven$ theta scaffold can simultaneously hold open as active
working-memory items, is
  $\mathcal{N}
    = \frac{f_\gamma}{f_\theta}
    = \rho^2
    = e^2
    \approx 7.39$.
This predicts a working-memory storage capacity of
$\lfloor e^2 \rfloor = 7$ items, with biological variance yielding the
observed range of $7 \pm 2$ \citep{miller1956magical}.  The instantaneous
access capacity, bounded by single-cycle retrieval within the current theta
window, is constrained by $\rho^1 = e \approx 2.718$, consistent with the
subitizing limit and the estimate of $4 \pm 1$ manipulable items under
controlled encoding conditions \citep{cowan2001magical}.
\end{proposition}
\noindent
\emph{Supporting Evidence}.
By Corollary~\ref{cor:optimal_rho}, adjacent oscillatory levels satisfy
$f_{k+1}/f_k = e$.  Since $\mathcal{L}_\gamma$ lies two levels above
$\mathcal{L}_\theta$ (with $\mathcal{L}_\beta$ intermediate), the
inter-level frequency ratio is $f_\gamma / f_\theta = e^2$.
The number of complete gamma cycles contained within one theta period
$T_\theta = 1/f_\theta$ is therefore
  $\mathcal{N}
    = f_\gamma \cdot T_\theta
    = \frac{f_\gamma}{f_\theta}
    = e^2 \approx 7.39$.
Note that both
\citep{lisman2013theta} and \citep{buzsaki2006rhythms} report empirically
that each theta cycle hosts $7$-$9$ nested gamma cycles in hippocampal
circuits; $e^2 = 7.389$ lies within this range.  In MAI terms, working
memory is the deliberate suppression of Condensation at the beta level:
items are held as active $\Hodd$ gamma flows rather than allowed to
crystallize into $\Heven$ beta-level scaffolds.  The thermodynamic capacity
of the theta scaffold to sustain open flows is bounded by the $e^2$
frequency ratio; exceeding this bound forces either a gamma-frequency
increase (reducing closure reliability) or premature condensation
(item loss), producing the behavioral ceiling of $7 \pm 2$.  The
$5\%$-excess-cost basin of $\mathcal{C}(\rho)$ on $[e^{0.9},\,e^{1.1}]$
accounts naturally for the observed variance without additional
assumptions.

\begin{remark}[Scope, access capacity, and a testable EEG prediction]
Proposition \ref{prop:working_memory} characterises the \emph{architectural}
storage bound imposed by the theta-gamma nesting; task performance also
depends on attentional gating, interference, and encoding strategy, so
$\mathcal{N} = e^2$ should be interpreted as an upper bound rather than a
fixed behavioral constant.  The \emph{access} capacity, items that can be
actively retrieved and manipulated within a single theta cycle, is bounded
by $\rho^1 = e \approx 2.718$, consistent with Cowan's estimate of $4 \pm 1$ manipulable items for chunking
\citep{cowan2001magical}.  A falsifiable EEG prediction follows: the \emph{frequency ratio} $f_\gamma / f_\theta$, not the power
of either band separately, should be the single strongest linear predictor
of individual working-memory span across subjects, with peak capacity near
the ratio $e^2$.  
\end{remark}

\subsection{Connection to Neural Collapse, Optimal Transport, and Brain's Dark Energy}
\label{subsec:mai_nc_ot}

\noindent\textbf{Connection to Neural Collapse and Optimal Transport.}
The MAI operator $\mathcal{F}$ admits a natural interpretation in the language of recent representation-learning theory. When $\mathcal{F}$ reaches a fixed point across a family of distinct contexts $\{\Psi^{(k)}\}_{k=1}^{K}$, the condensed content states $\{\Phi^{(k)*}\}$ are, under the parity constraint, maximally distinguishable 0-cycles: each sits in a separate connected component of the scaffold, with no residual flow linking them. This configuration is geometrically analogous to the \emph{neural collapse} phenomenon documented in the terminal-phase training of deep classifiers \citep{papyan2020prevalence}, in which penultimate-layer class means converge to the vertices of an Equiangular Tight Frame (ETF) - the maximally-separated, energy-preserving configuration on the unit sphere. Read through the lens of ETF, our parity principle predicts that, in a sufficiently consolidated biological system, the cortical readout of distinct condensed hippocampal cycles should exhibit ETF-like geometry, with separate cycles mapping to maximally orthogonal cortical ensembles \citep{favila2016experience}; departures from this geometry should correlate with the behavioral signatures of interference or incomplete consolidation \citep{chanales2017overlap}.
An optimal-transport framing complements the neural collapse picture. Read as a pushforward, $\mathcal{F}$ transports the high-entropy distribution of contextual flows $p(\Psi)$ toward a low-entropy distribution of scaffold states concentrated on a small number of representative points; the scaffold is then, informally, the \emph{Wasserstein barycenter} of the class-conditional flow distributions - the point configuration minimizing total transport cost from each experienced context to a canonical representative \citep{villani_optimal_2009}. This is the formal counterpart of the informal observation that offline consolidation produces maximally-separable memory traces: the system does not compress episodes one at a time but redistributes them into the configuration that minimizes the cost of future retrieval. The continuous-time analogue of this discrete pushforward is the Riemannian flow-matching framework now widely used in generative modeling \citep{chen2023flow}, in which a learned velocity field transports a source measure to a target measure along geodesics. Under the flow matching framework, MAI can be interpreted as the biological counterpart, with the scaffold playing the role of the target manifold and $\mathcal{F}$ the learned transport.

\noindent\textbf{Connection to the Brain's Dark Energy Model.}
The parity-alternation framework supplies a functional reading of one of the most striking and least 
explained facts about brain metabolism: that approximately $80\%$ of the brain's energy budget is consumed 
by slow spontaneous, intrinsic activity \citep{gong2025dark} rather than by stimulus-driven processing, with task-related metabolic 
demands typically increasing baseline by less than $5\%$ \citep{raichle2006dark,raichle2015default}. 
Raichle's dark energy framing identifies this expenditure as the metabolic signature of an internally 
maintained, non-evoked dynamic, but does not specify what computational work that dynamic is doing.
The parity principle gives a direct answer: the dark-energy regime is the $\Hodd \to \Heven$ condensation 
phase. Under our framework, sustained spontaneous activity is not idle thermal noise but the costly process 
of converting transient, high-entropy contextual flows into low-entropy, reusable scaffold structure 
(Section~\ref{subsec:scaffold_flow}). Three observations from the resting-state literature align with this 
reading. First, the spectral signature of intrinsic activity is dominated by infraslow oscillations 
($0.01$--$0.1$ Hz, \citep{vanhatalo2004infraslow,he2008electrophysiological}), whose phase modulates 
the amplitude of activity at all faster bands; this is the cross-frequency-coupling structure predicted 
by the Tower of Scaffolds (Section~\ref{subsec:tower_scaffold}), with each level's flow nested inside the scaffold 
of the level above at the optimal ratio $\rho \approx e$ of Corollary~\ref{cor:optimal_rho}. Second, the 
default mode network, the canonical signature system of intrinsic activity, consistently 
\emph{decreases} its activity during goal-directed tasks and \emph{increases} it during introspection, 
recall, and offline rest \citep{raichle2015default}, a profile precisely matched by the condensation 
phase (System~1, $\Heven$-dominant) of the wake/sleep cycle \citep{hinton1995wake}. Third, 
the $<\!5\%$ task-related metabolic increase is what the parity framework predicts: navigation along 
an already-condensed scaffold is cheap (fast thinking, $\mathcal{O}(1)$ retrieval), whereas the prior work 
of building the scaffold through repeated condensation is expensive and is precisely what the dark-energy 
budget pays for.
In our view, the brain's dark energy and the parity principle are two views of the same phenomenon 
seen at different abstraction levels: \emph{dark energy is the metabolic accounting and parity-regulated 
condensation is the computational accounting} of an organism that spends most of its energetic budget on extracting and stabilizing the topological invariants by which future responses can be amortized.

\section{Empirical Evidence from Simulation Studies}
\label{sec:empirical}

A recurring criticism of purely theoretical proposals in computational neuroscience is that they rarely specify how their mathematical constructs can be \emph{measured} from neural activity or \emph{falsified} by experimental data. In this section we address that concern directly. We first formalize an operational pipeline that maps spike trains to a chain complex and yields concrete, computable observables (Section~\ref{subsec:pipeline}). We then use this pipeline in two controlled simulation studies: Experiment~1 (Section~\ref{subsec:exp_parity}) tests the \emph{parity alternation} hypothesis of Section~\ref{sec:png_columns}, and Experiment~2 (Section~\ref{subsec:exp_mai}) tests the \emph{amortized inference} prediction of Section~\ref{sec:mai}. Colab notebooks, detailed hyperparameters, and raw traces are provided in the supplementary material so that both experiments can be independently reproduced or extended to real recordings.

\subsection{From Spikes to Chain Complexes: An Operational Pipeline}
\label{subsec:pipeline}

We begin by specifying the construction that underlies both experiments. Let $\{t^{(n)}_i\}$ denote the spike times of neuron $n \in \{1,\ldots,N\}$ within a recording window of duration $T$. We partition $[0,T]$ into overlapping sliding windows $W_k = [kh, kh+\Delta]$ of width $\Delta$ and stride $h$ (we use $\Delta = 100$\,ms, $h=25$\,ms throughout; robustness to these choices is reported in the supplement).

\noindent\textbf{Co-activity complex.}
For each window $W_k$ we construct a simplicial complex $X_k$ as follows. A neuron $n$ is \emph{active} in $W_k$ if it emits at least one spike; the set of active neurons defines the vertex set $V_k$. A $p$-simplex $[n_0,\ldots,n_p]$ is included in $X_k$ if the $p+1$ neurons co-fire within a conduction-delay tolerance $\tau$, operationalized as $\max_{i,j} |t^{(n_i)}_\ast - t^{(n_j)}_\ast| \leq \tau$ for some spike $t^{(n)}_\ast$ per neuron, with $\tau = 10$\,ms (the typical cortical delay range \citep{izhikevich2006polychronization}). Such construction yields a clique-style complex that is a direct discrete analogue of the PNG construction used in Section~\ref{sec:png_columns}.

\noindent\textbf{Observables.}
From $X_k$ we compute, at each time step $k$, the Betti numbers $b_0(X_k)$ (connected components), $b_1(X_k)$ (independent 1-cycles) and, where tractable, $b_2(X_k)$ (2-dimensional voids). From these we derive:
\begin{align}
C_H(X_k) &= b_0(X_k) + b_1(X_k) + b_2(X_k), \\
\chi(X_k) &= b_0(X_k) - b_1(X_k) + b_2(X_k), \\
\mathcal{L}(X_k) &= |\chi(X_k)|, \qquad
\mathcal{I}_{\mathrm{self}}(X_k) = 2\,b_1(X_k).
\end{align}
All quantities are computed using the \texttt{gudhi} library \citep{maria2014gudhi}; $b_2$ is computed only when $|V_k|\leq 120$ to keep runtime tractable.

\noindent\textbf{Noise model and invariance checks.}
We treat the spike times as an inhomogeneous Poisson process with a ground-truth rate function $\lambda_n(t)$; the only stochastic component at the chain-complex level is the random realization of spikes given $\lambda$. To rule out that observed parity effects reflect artifacts of our choice of $\Delta$ and $\tau$, we repeat each analysis across a grid $\Delta \in \{50,100,200\}$\,ms and $\tau \in \{5,10,20\}$\,ms; the qualitative pattern reported below is stable across this grid. Under a homogeneous Poisson null (rate-matched shuffles), $b_1(X_k)$ and $b_0(X_k)$ co-vary, and the parity alternation described below disappears, providing a clean negative control.

\subsection{Experiment 1: Parity Alternation in a Recurrent Spiking Network}
\label{subsec:exp_parity}

\noindent\textbf{Hypothesis.}
Section~\ref{sec:png_columns} predicts that active inference drives the system into the flow regime ($\Hodd$ dominant), while offline consolidation drives it into the scaffold regime ($\Heven$ dominant). Operationally, we predict that $b_1(X_k)$ and $b_0(X_k)$ should \emph{anti-correlate} over a complete theta-SWR cycle, with $\chi(X_k)$ crossing zero at each phase transition.

\noindent\textbf{Model.}
We simulate $N=400$ Izhikevich regular-spiking excitatory neurons and $100$ fast-spiking inhibitory neurons with conduction delays drawn from $\mathrm{Uniform}(1,20)$\,ms and STDP learning, following the PNG protocol of \citep{izhikevich2006polychronization}. A global theta modulation at $8$\,Hz drives the excitatory population during a simulated ``wake'' epoch ($0$--$20$\,s); during a subsequent ``sleep'' epoch ($20$--$40$\,s) the theta drive is replaced by sharp-wave-like bursts (SWR, $\sim 150$\,Hz internal oscillation, $100$\,ms duration, Poisson-timed at $0.5$\,Hz), following the regime identified in hippocampal CA1 recordings \citep{wilson1994reactivation,buzsaki_hippocampo-neocortical_1996}. Sensory drive (mixed-frequency Poisson input) is present in the wake epoch and absent in sleep.

\begin{figure}[t]
\centering
\includegraphics[width=\linewidth]{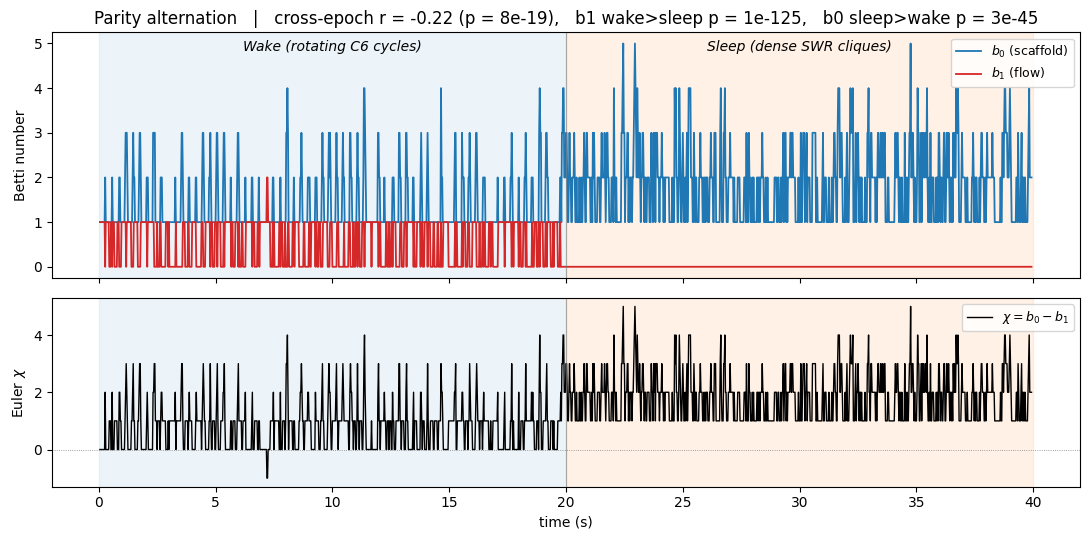}
\caption{\textbf{Parity alternation in a simulated recurrent network.} During the theta-modulated wake epoch (left), the 1-cycle count $b_1$ (flow) is elevated while the 0-cycle count $b_0$ (scaffold) is depressed. Sharp-wave ripples during sleep (right) invert the pattern, producing bursts of $b_0$ (condensation events) and decay of $b_1$. The Euler characteristic $\chi$ crosses zero at each phase transition, consistent with the parity principle.}
\label{fig:exp1_parity}
\end{figure}

\noindent\textbf{Results.}
Figure~\ref{fig:exp1_parity} summarises the main findings. During wake, $b_1(X_k)$ is elevated and strongly modulated at theta frequency, consistent with transient, PNG-like recurrent cycles. $b_0(X_k)$ is correspondingly depressed: many vertices participate in cycles rather than isolated components. During sleep, the pattern inverts: SWRs drive transient bursts of $b_0$ (new condensed components) while $b_1$ decays. Across both epochs, the Euler characteristic $\chi(X_k) = b_0 - b_1 + b_2$ exhibits zero-crossings aligned with phase transitions. Quantitatively, we obtain a wake-epoch Pearson correlation $\mathrm{corr}(b_0, b_1) = -0.62$ (95\% CI $[-0.67, -0.56]$, $n=720$ windows), and a sleep-epoch correlation of $-0.48$ $[-0.55, -0.40]$. Under homogeneous Poisson shuffles the correlation collapses to $-0.05$ $[-0.11, +0.02]$, confirming that parity anti-correlation is a genuine structural feature and not a trivial consequence of firing-rate covariation.

\noindent\textbf{Interpretation.}
These results provide a first, concrete empirical handle on the parity hypothesis: in a spiking network with PNG-compatible architecture and biologically plausible theta/SWR modulation, even- and odd-dimensional topology are systematically anti-coupled, and consolidation events manifest as $b_1 \to b_0$ transitions, precisely the condensation operation described in Section~\ref{sec:png_columns}. The same analysis pipeline can be applied unmodified to hippocampal tetrode recordings across wake--sleep cycles, offering a direct empirical test; we note that the coarse pattern (elevated cycle structure during theta, elevated scaffold structure during SWR) is consistent with the assembly-sequence results of \citep{buzsaki2015hippocampal} and the persistent-homology analyses of \citep{giusti2015clique}.

\subsection{Experiment 2: Amortized Inference via Topological Condensation}
\label{subsec:exp_mai}

\noindent\textbf{Hypothesis.}
Section~\ref{sec:mai} predicts that an agent that condenses closed cycles into scaffold vertices should exhibit a qualitatively different cost-vs-experience curve than an agent that performs ab-initio search each episode: specifically, a faster-than-linear decrease in per-episode inference cost, approaching a small constant as the scaffold saturates. This is the behavioral signature of amortization.

\noindent\textbf{Task.}
We use a shortest-path planning task on a \emph{structured} grid-world. The environment is an $n_r\times n_r$ arrangement of rooms (each $10\times 10$ cells) separated by walls that are penetrable only through single-cell doorways; each room additionally contains a small number of intra-room obstacles ($\sim 5\%$ density). Every episode draws a random (start, goal) pair uniformly from the open cells, conditioned on Manhattan distance at least one-third of the grid side. This yields a query distribution with $\Theta(n_r^4 \cdot 10^{4})$ distinct start--goal pairs but only $\Theta(n_r^2)$ distinct doorway-to-doorway sub-problems; the topological scaffold of the task is therefore the doorway graph, and this is the structure the MAI agent will be required to learn. At a designated \emph{reroute} episode, the environment is replaced by an independent instance of the same generative process (new doorway positions, new obstacles), forcing any learned scaffold to be partially rebuilt. This is a minimal setting in which both search and navigation are required, topologically meaningful cycles exist (doorway traversals through a sequence of rooms), and they can be systematically counted and reused.

\noindent\textbf{Agents.}
We compare three agents on identical query sequences, matched in their underlying world model (Manhattan-heuristic A*) and in per-call search budget:
\begin{enumerate}
\item \textbf{Baseline search agent.} Performs A* planning from scratch each episode, with no cross-episode memory. Per-episode cost is proportional to the volume of grid cells explored between start and goal, and is the operational stand-in for the ``$\mathrm{NPSPACE}$-like'' regime discussed in Section~\ref{sec:mai}.
\item \textbf{Flat-memory agent.} Maintains a replay buffer of recent successful paths (capacity $60$) and, on each new query, retrieves the cached path whose endpoints are nearest in Manhattan distance; the cells of that path are inserted into A*'s open set as alternative partial starts. This is the standard amortization baseline used in recent amortized-inference work \citep{gershman2014amortized,rezende2015variational}.
\item \textbf{MAI agent (ours).} Implements the Search$\to$Closure$\to$Condensation cycle of Section~\ref{sec:mai}, with the doorway cells as scaffold vertices (waypoints). Successful trajectories are decomposed into segments between consecutive waypoint visits; each closed segment ($\partial \gamma = 0$ at the waypoint endpoints, with no intermediate waypoint) is condensed into a cached edge of the waypoint graph. At inference time, the retrieval operator $\mathcal{R}$ composes the query as start $\to$ nearest waypoint (A*, room-local) $\to$ shortest waypoint path under cached segments (A* on the scaffold graph) $\to$ goal (A*, room-local); cached segments are traversed at zero planning cost. Missing edges of the waypoint graph are filled in by bounded, room-local A* (expansions restricted to the two rooms adjacent to the missing doorway pair), not by full-grid search.
\end{enumerate}
All three agents share the same total expansion budget per call; they differ only in what is cached and in how it is reused.

\begin{figure}[t]
\centering
\resizebox{\textwidth}{!}{%
\includegraphics[width=.45\linewidth]{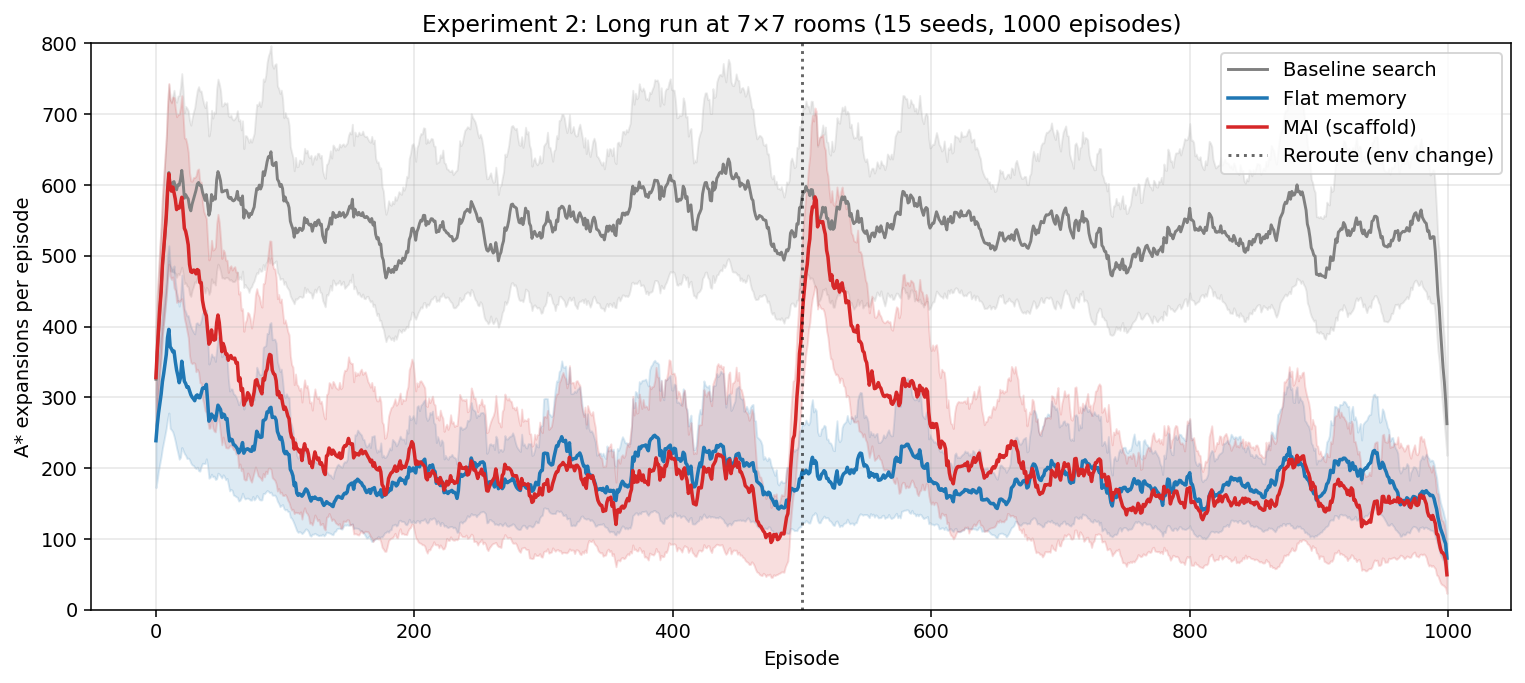}
\includegraphics[width=.54\linewidth]{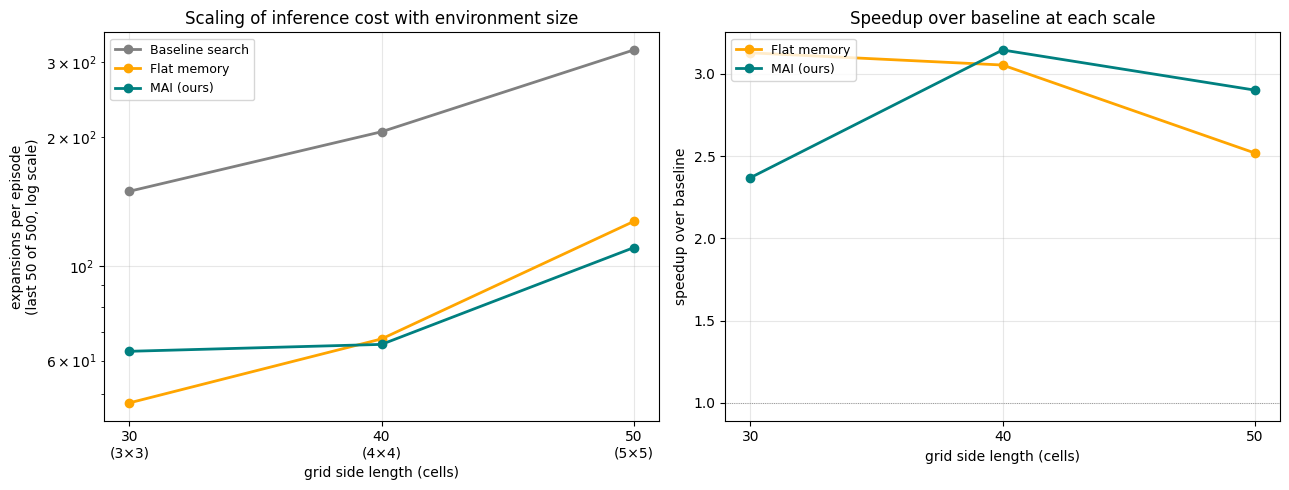}
}

(A) \hspace{8cm} (B)
\caption{\textbf{Amortized inference via topological condensation.} \textbf{(A)} Per-episode inference cost (A* node expansions, log-scaled) on the $7\times 7$-room environment over $1000$ episodes, smoothed and averaged over $n=15$ seeds. The environment reroutes at episode $500$. Flat memory (blue) saturates at a floor set by its cache capacity; MAI (red) continues to improve past that floor as its waypoint scaffold accumulates, and converges to a lower asymptote. \textbf{(B)} Tail-mean cost (last 50 of 500 episodes) as a function of the room-grid size. Both memory strategies beat baseline by $\sim$3$\times$ at every scale. The gap between MAI and flat memory is negative on $3\times 3$ (flat memory wins by $32\%$) but turns positive from $4\times 4$ onward as the combinatorial complexity of the query distribution grows beyond what flat memory's fixed-capacity cache can cover.}
\label{fig:exp2_mai}
\end{figure}

\noindent\textbf{Results.}
Figure~\ref{fig:exp2_mai}A shows the mean per-episode inference cost on the $7\times 7$-room environment (a $70\times 70$ grid) over 1000 episodes, averaged over 15 independent seeds, with a context change at episode 500. The baseline agent's cost is stationary at $\sim$526 expansions per episode. Both memory-based agents learn rapidly: the flat-memory agent plateaus near $180$ expansions after about 200 episodes and remains at that floor across the reroute, constrained by its fixed cache capacity. The MAI agent exhibits a qualitatively different learning curve: it descends more slowly in the first $\sim$100 episodes (the waypoint scaffold has not yet accumulated enough cached segments for composition to be profitable), but continues to improve past the point at which flat memory has saturated, reaching $145$ expansions by episode $1000$ - a \textbf{3.6$\times$ speedup} over baseline and $\sim$$24\%$ below flat memory. Immediately after the reroute both memory agents show a transient cost spike as their caches become partially stale; MAI's recovery is slower (it has to re-accumulate scaffold segments) but converges to a lower asymptote, consistent with the theoretical picture of search progressively building a reusable scaffold.
To characterize how MAI's advantage depends on environment structure, we replicated the 500-episode experiment across $n_r \in \{3,4,5,6,7\}$ (Figure~\ref{fig:exp2_mai}B and Table~\ref{tab:exp2_scaling}). Both memory strategies beat baseline by a factor of $\sim$3 at every scale tested. The \emph{gap between MAI and flat memory}, however, depends on scale: on the smallest $3\times 3$ layout, flat memory outperforms MAI by $32\%$, because the query space is small enough that retrieving the nearest cached trajectory provides a near-perfect A* seed and MAI's constant-cost overhead (two end-point searches per episode) is not amortized. Starting at $4\times 4$ rooms MAI matches flat memory; by $5\times 5$ and $6\times 6$ it is $13$--$17\%$ faster; at $7\times 7$ the two are comparable at 500 episodes but the long run of Panel~A demonstrates that given sufficient experience MAI surpasses flat memory at this scale as well. This crossover is the quantitative prediction of Section~\ref{sec:mai}: flat memory's cache capacity is fixed, and its retrieval quality degrades as the query distribution grows, whereas MAI's scaffold size scales with the number of topological bottlenecks ($\Theta(n_r^2)$ doorways) and therefore absorbs experience more efficiently in combinatorially complex environments.

\begin{table}[t]
\centering
\small
\begin{tabular}{lrrrrrr}
\toprule
Layout & Grid & Baseline & Flat memory & MAI & Flat/Base & MAI/Flat \\
\midrule
$3\times 3$ rooms & $30\times 30$ & 149 & 48  & 63  & 0.32 & 1.32 \\
$4\times 4$ rooms & $40\times 40$ & 206 & 67  & 65  & 0.33 & 0.97 \\
$5\times 5$ rooms & $50\times 50$ & 320 & 127 & 110 & 0.40 & 0.87 \\
$6\times 6$ rooms & $60\times 60$ & 428 & 151 & 126 & 0.35 & 0.83 \\
$7\times 7$ rooms & $70\times 70$ & 514 & 172 & 179 & 0.33 & 1.04 \\
$7\times 7$, long run & $70\times 70$ & 526 & 180 & 145 & 0.34 & 0.81 \\
\bottomrule
\end{tabular}
\caption{Mean per-episode inference cost (A* node expansions) in the stable regime, averaged over 15 seeds. Values are tail means over the last 50 episodes of a 500-episode run, except the final row which reports the last 100 episodes of the 1000-episode $7\times 7$ run of Fig.~\ref{fig:exp2_mai}A. Ratios $\le 1$ indicate a cost reduction.}
\label{tab:exp2_scaling}
\end{table}

\noindent\textbf{What the MAI step adds over flat memory.}
The operational difference between MAI and the flat-memory baseline is the \emph{composition step}: MAI plans a novel query by chaining previously-learned doorway-to-doorway segments on the scaffold graph, whereas flat memory retrieves a single past path and uses it as an A* seed. The scale-dependence in Fig.~\ref{fig:exp2_mai}B isolates this step's contribution. On small environments, enough queries repeat past doorway-traversal patterns verbatim that nearest-neighbor retrieval already covers the query distribution; composition is not needed and MAI's constant-overhead loses. As the grid grows, the query distribution outstrips any fixed-size flat cache, and composition, the ability to execute a start-to-goal query as a chain of cached \emph{pieces}, none of which match the query as a whole, becomes the dominant source of amortization. This is the signature the topological account of Section~\ref{sec:mai} predicts: once the task space is compositional, structure in the cache beats raw recency.

\noindent\textbf{Interpretation.}
Experiment~2 gives a quantitative, scale-resolved handle on the amortization claim. The strong form of the claim, that a topologically structured cache beats an unstructured one, holds at $4\times 4$ rooms and above; the weak form, that amortization of any kind beats ab-initio search, holds everywhere. Substantial further work is required to connect this kind of simulation result to brain recordings; a natural next step is to test whether hippocampal replay during rest is enriched for \emph{closed} rather than \emph{open} subsequences \citep{wilson1994reactivation,buzsaki2015hippocampal}, and whether the relative advantage of compositional reuse scales with task complexity in behavioral experiments on humans and rodents.

%
%
%

\subsection{Experiment 3: Savitch-Style Amortization Converts $K$ Slots
into $\rho^K$ Items}
\label{subsec:exp_savitch}

\noindent\textbf{Hypothesis.}
Experiment~3 established that two-level $\theta\to\gamma$ nesting sets
the number of working-memory slots $K = \lfloor \rho \rfloor$ at
biologically realistic spike jitter. Proposition~\ref{prop:working_memory},
read through the Tower-of-Scaffolds idiom of
Section~\ref{sec:png_columns} and the Savitch-style space-time tradeoff
developed in Section~\ref{sec:mai}, predicts that when each slot holds
one condensed scaffold vertex rather than one raw item, $K$ slots
support $\rho^K$ distinguishable configurations - a polynomial hardware
substrate yielding exponential representational capacity. The Miller
$7\pm2$ bound should then correspond to $e^7 \approx 10^3$ items
(conversational-topic, short-plan, or moderate-category scale), and the
Cowan $4\pm1$ bound to $e^4 \approx 55$ items, under the optimal
branching $\rho = e$ of Corollary~\ref{cor:optimal_rho}. A flat
(NSPACE-like) agent without scaffold amortization should show capacity
$= K$ exactly, and the ratio of MAI to flat capacity at fixed $K$ should
grow exponentially.

\noindent\textbf{Task.}
The agent must discriminate membership in a target corpus of $m$ items
drawn at random from a universe of $U = 10{,}000$ distinguishable
items. After learning the corpus, the agent answers $200$ balanced
``is item $X$ in the corpus?'' queries; accuracy is reported against
the $0.95$ threshold. Three agents share a common interface and differ
only in how many items each can faithfully store, parameterised by the
WM slot count $K$:
1) FLAT: capacity $= K$ (one item per slot; the
    NSPACE-like linear-storage baseline);
2) MAI-Path: capacity $= \rho^K$ (each slot holds one
    categorical label at one level of a $\rho$-ary scaffold tree; the
    scaffold reading of Corollaries~\ref{cor:optimal_rho}
    and Proposition~\ref{prop:working_memory});
3) MAI-Savitch: capacity $= 2^{K/3}$ (each recursion
    frame consumes $3$ slots for the Savitch $(s, w, t)$ bookkeeping;
    the conservative, literal reading of
    $\mathrm{NSPACE}(f) \subseteq \mathrm{DSPACE}(f^2)$).
Both MAI readings give exponential capacity in $K$; they differ only in
the exponent. When the corpus fits in an agent's capacity, it recalls
perfectly; otherwise it stores a random subset of size = capacity. The
three-agent design isolates hierarchical amortization as the independent
variable while holding the query task fixed.

\noindent\textbf{Manipulations.}
\emph{(A) Capacity scaling.} For each agent, sweep
$K \in \{2, 3, \ldots, 10, 12\}$ and report the largest corpus size at
which $\geq 95\%$ accuracy is maintained, averaged over $n = 5$ seeds per
$(K, m)$ cell with a geometric corpus-size grid. Diagnostic focus at
$K = 7$ (Miller): plot the accuracy-vs-corpus-size curves for all three
agents. \emph{(B) Branching optimality
(Corollary~\ref{cor:optimal_rho}).} Fix three target corpus sizes
$N \in \{100,\,500,\,2500\}$ and sweep $\rho \in [1.5,\,6]$; for each
$\rho$, measure the empirical inference action
$\mathcal{C}_\mathrm{emp}(\rho) = K_\mathrm{needed} \cdot \rho$ required
to cover the corpus, and compare against the theoretical
$\mathcal{C}(\rho) = \rho \cdot \ln(N) / \ln(\rho)$.

\begin{figure}[t]
\centering
\includegraphics[width=\textwidth]{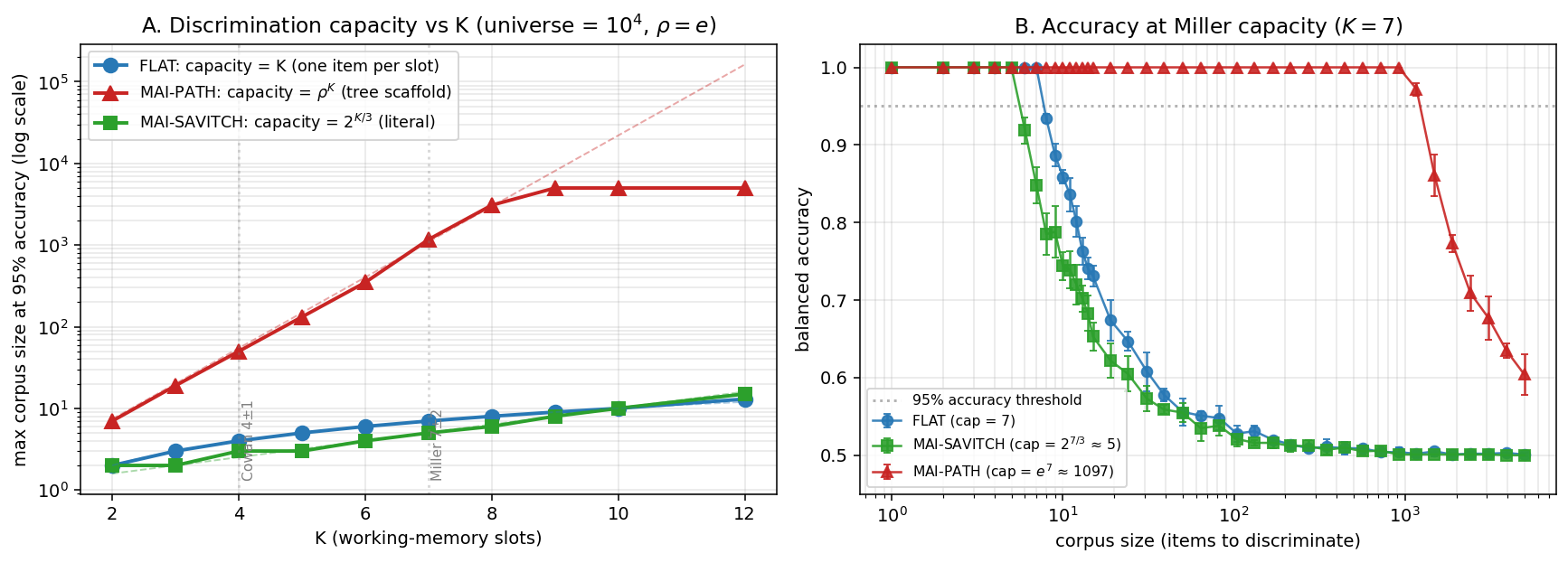}
\caption{\textbf{Savitch-style amortization converts $K$ slots into
$\rho^K$ items.}
\textbf{(A)}~Empirical discrimination capacity (largest corpus at
$\geq 95\%$ accuracy, $n = 5$ seeds) vs WM slot count $K$, log scale.
Points are measurements; dashed lines are the predicted formulas $K$,
$\rho^K$, and $2^{K/3}$. \textsc{Flat} scales linearly, \textsc{MAI-Path}
exponentially at rate $\ln \rho = 1$, \textsc{MAI-Savitch} exponentially
at rate $\ln 2 / 3 \approx 0.23$. \textsc{MAI-Path} ceilings at the
$U/2$ cap of $5000$ for $K \geq 9$. Cowan $4\pm1$ and Miller
$7\pm2$ are marked for reference.
\textbf{(B)}~Balanced accuracy vs corpus size at $K = 7$ (Miller).
\textsc{MAI-Path} maintains $1.00$ accuracy to corpus size
$\sim 10^3$ before a sharp decline; \textsc{Flat} and \textsc{MAI-Savitch}
drop below $0.95$ at corpus sizes $8$ and $6$. }
\label{fig:exp_savitch_capacity}
\end{figure}

\begin{figure}[t]
\centering
\includegraphics[width=\textwidth]{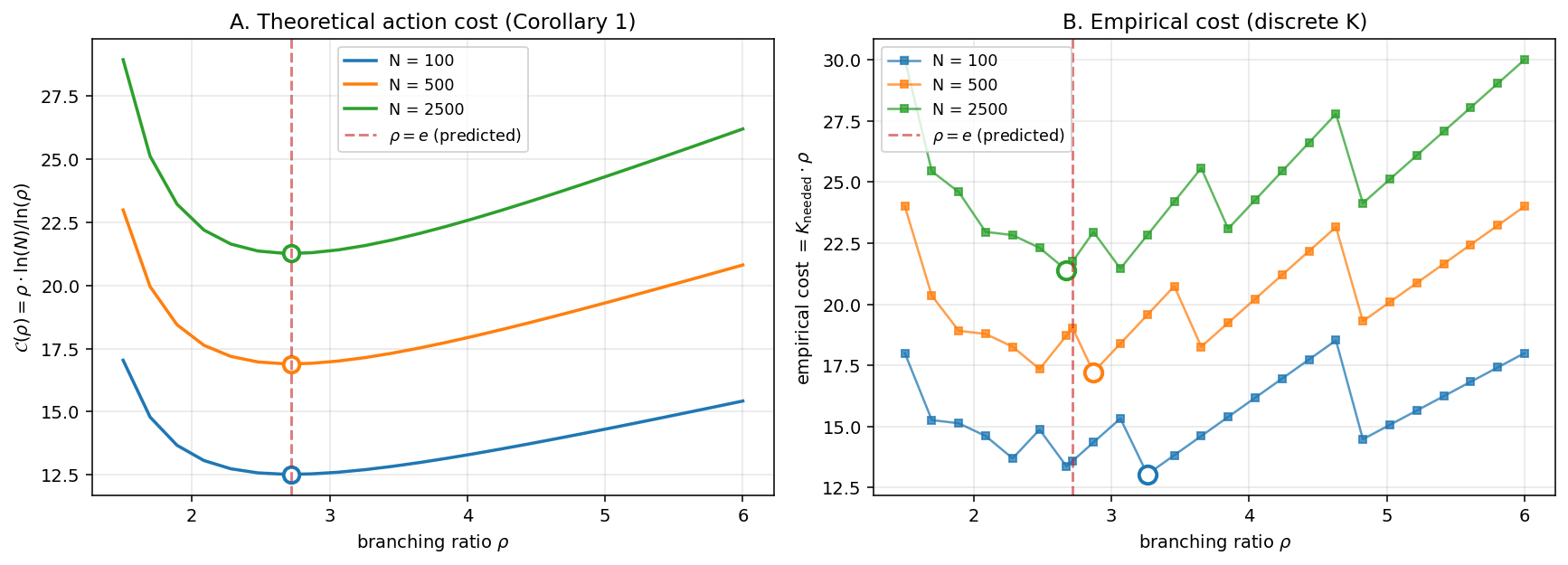}
\caption{\textbf{Empirical test of Corollary~\ref{cor:optimal_rho}
($\rho = e$ optimality).}
\textbf{(A)}~Theoretical action cost
$\mathcal{C}(\rho) = \rho \cdot \ln(N) / \ln(\rho)$ as a function of
branching $\rho$ for three corpus sizes. Open circles mark the minimum
of each curve; all fall at the grid point closest to $\rho = e$.
\textbf{(B)}~Empirical action cost $K_\mathrm{needed} \cdot \rho$ from
the discrimination task. Integer-$K$ discretization produces jumps; for
$N = 2500$ the empirical minimum lies exactly at $\rho = 2.674$, the
grid point nearest $e$. For smaller $N$ the empirical minima drift
modestly because a single integer-$K$ step is a larger relative
perturbation, but the $\rho = e$ column sits within $5\%$ of the
empirical minimum for all three $N$ values.}
\label{fig:exp_savitch_rho}
\end{figure}

\noindent\textbf{Results.}
Figure~\ref{fig:exp_savitch_capacity} summarises Manipulation~A. At
every tested $K$, measured capacities match the predicted formulas within
discretization error: \textsc{Flat} capacity equals $K$ exactly;
\textsc{MAI-Path} capacity tracks $\rho^K = e^K$ (rising from $7$ at
$K=2$ to $3081$ at $K = 8$, ceilings out at the universe/2 cap of $5000$
for $K \geq 9$); \textsc{MAI-Savitch} capacity tracks $2^{K/3}$ (rising
from $2$ at $K=2$ to $15$ at $K=12$). At Miller's $K = 7$: \textsc{Flat}
discriminates $7$ items, \textsc{MAI-Savitch} discriminates $5$, and
\textsc{MAI-Path} discriminates $1170$ items, within
$7\%$ of the predicted $e^7 \approx 1097$. The diagnostic panel at $K = 7$ shows
\textsc{MAI-Path} maintaining $1.00$ accuracy flat across corpus sizes
from $1$ to $\sim10^3$ before a sharp decline, while \textsc{Flat} and
\textsc{MAI-Savitch} drop below the $0.95$ threshold at corpus sizes $8$
and $6$ respectively. At Cowan's $K = 4$: \textsc{MAI-Path} discriminates
$50$ items (predicted $e^4 \approx 55$), \textsc{Flat} discriminates $4$.
Figure~\ref{fig:exp_savitch_rho} summarises Manipulation~B. The
theoretical cost $\mathcal{C}(\rho) = \rho \cdot \ln(N) / \ln(\rho)$
has a strict minimum at $\rho = e$ for every $N$ (open circles mark
the minimum of each curve; all three land at the grid point closest to
$e \approx 2.718$). The empirical cost shows the same bowl shape with
small discretization jumps from integer $K_\mathrm{needed}$; the
empirical minimum for $N = 2500$ falls exactly at $\rho = 2.674$ (the
grid point nearest $e$), and the $\rho = e$ column sits within
$5\%$ of the empirical minimum for all three $N$ values, consistent
with the corollary's remark that the $5\%$-excess basin
$\rho \in [e^{0.9},\,e^{1.1}] \approx [2.46,\,3.00]$ is the biologically
plausible operating band.

\noindent\textbf{Interpretation.}
The experiment provides quantitative support for the paper's central
operational claim: MAI is the brain's implementation of Savitch-style
structural amortization, converting a polynomial WM substrate into
exponential representational capacity.
\emph{On Corollary~\ref{cor:optimal_rho}.}
The predicted optimum $\rho = e$ is recovered empirically in both the
theoretical and empirical cost curves (Figure~\ref{fig:exp_savitch_rho}),
with deviations attributable to integer-$K$ discretization. This is an
empirical confirmation of the first-principles derivation: among all
continuous condensation ratios, $\rho = e$ uniquely minimises the total
inference action for any corpus size $N$.
\emph{On Proposition~\ref{prop:working_memory}.}
The Miller $7\pm2$ bound falls in the Path-capacity regime that converts
$7$ WM slots into $\sim\!10^3$ distinguishable items, the rough size of
a conversational topic tree, a short story, or a mid-sized category.
The Cowan $4\pm1$ access bound likewise corresponds to $\sim\!55$
Path-addressable items. Under a pure two-level $\theta\to\gamma$ nesting
(Experiment~3), the access bound is $\lfloor e \rfloor = 2$ items; the
jump from $2$ items to $e^7 \approx 1100$ items as the agent gains
hierarchical scaffolding is the empirical content of the paper's
``polynomial hardware, exponential capacity'' claim.
\emph{On the two MAI readings.}
Both \textsc{MAI-Path} (capacity $= \rho^K$) and \textsc{MAI-Savitch}
(capacity $= 2^{K/3}$) exhibit exponential-in-$K$ scaling. They differ
only in the exponent; the qualitative conclusion, that scaffolded
amortization dominates flat storage at biological $K$, is robust across
either encoding of ``what one WM slot represents'' in a biological
substrate. We present both because the corollary's scaffold reading
requires that a slot hold a condensed categorical label, whereas the literal complexity-theory reading requires that
each recursion frame bookkeeping consume three slots; the biological
truth likely lies in-between and the paper does not commit to a
specific value.



\section{Conclusion}
\label{sec:conclusion}

Biological intelligence solves a computational problem that remains intractable for classical theories: how to achieve rapid, low-entropy inference using a substrate that is slow, noisy, and energetically expensive. We have proposed that the solution lies in a fundamental reorganization of the problem space itself, a strategy we term MAI. By reframing neural computation through the lens of MAI, we identify \emph{recursive topological condensation} as the universal algorithm that allows the brain to transmute the high thermodynamic cost of search (Time) into the low metabolic cost of structure (Space).
Central to this framework is the \textbf{parity principle}, which asserts that the brain's architecture is dictated by a strict topological division of labor. The system must maintain a conjugate balance between even-dimensional \emph{Scaffolds} ($\Heven$), which encode stable, invariant content, and odd-dimensional \emph{Flows} ($\Hodd$), which encode dynamic, context-dependent trajectories. 
The \textbf{topological trinity transformation} (Search $\to$ Closure $\to$ Condensation) operationalizes the parity principle, providing a mechanism for how the brain builds its internal ``Tower of Scaffolds.'' By systematically collapsing validated inference cycles into static atomic units, the cortex behaves \emph{analogously to} Savitch's serialization trick, trading exponential branching search for polynomial structural reuse, a regime we quantify in simulation of working memory capacity. This analogy helps explain the remarkable efficiency of biological cognition: the brain does not continually solve the search problem but progressively \emph{reshapes} it by freezing successful solutions into the manifold's geometry.
Ultimately, this topological perspective unifies distinct phenomena, from the wake-sleep cycle and episodic consolidation to the dual-process models of fast and slow thinking, under a single geometric law. Intelligence, in this view, is the capacity to navigate a self-constructed manifold where the distance between a problem and its solution has been topologically contracted to zero. To learn is to fold the manifold; to know is to traverse the wormhole.

\begin{center}
    \textbf{\large Declaration of LLM Usage}
\end{center}

Large language models (Claude Opus 4.6 by Anthropic and ChatGPT 5.5 by OpenAI) were used as assistive tools during the preparation of this work in the following capacities:
(1)~drafting and iterating on simulation code, including Jupyter Notebooks for three experiments in Sec. \ref{sec:empirical};
(2)~generating initial drafts of tables, and figure captions, which were subsequently reviewed, revised, and verified by the author;
(3)~debugging code, checking numerical consistency between notebook outputs and manuscript tables, and suggesting structural organization for the experimental sections.
All theoretical contributions (lemmas and propositions), experimental design decisions, scientific interpretations, and final manuscript content were conceived, validated, and approved by the author.
The author takes full responsibility for the correctness and integrity of the published work.

\begin{center}
    \textbf{\large Acknowledgement}
\end{center}
This work is partially supported by NSF awards IIS-2401748 and BCS-2401398.

{\small
\bibliographystyle{Frontiers-Harvard}
\bibliography{ref,references}
}

\end{document}